%% file: Fawaz-ArXiv-ITtrans-20090529.tex
\documentclass[11pt,onecolumn,draftclsnofoot]{./sty/IEEEtran}

\usepackage{graphicx,epsfig,psfrag,subfigure,color}
\usepackage{amsmath,dsfont,amssymb}
\usepackage{cite,url,balance}
\usepackage{array}

\input{sty/macros}

\begin{document}

\title{Asymptotic Capacity and Optimal Precoding in MIMO Multi-Hop Relay Networks}

\author{Nadia~Fawaz,     
        Keyvan~Zarifi,   
        Merouane~Debbah, 
        David Gesbert    
\thanks{Manuscript submitted to IEEE Transactions on Information Theory.
        This work was supported by the French Defense Body DGA, by BIONETS project (FP6-027748, www.bionets.eu) and by Alcatel-Lucent within the Alcatel-Lucent Chair on flexible radio at SUPELEC. The material in this paper was presented in part at IEEE Workshop on Information Theory (ITW 2008), Porto, Portugal, May 2008.}
\thanks{N. Fawaz and D. Gesbert are with the Mobile Communications Department, EURECOM, Sophia-Antipolis, France (email: \{nadia.fawaz,david.gesbert\}@eurecom.fr). K. Zarifi is with INRS-EMT \& Concordia University, Montr\'{e}al, Canada (email: keyvan.zarifi@emt.inrs.ca).
M. Debbah is with Alcatel-Lucent Chair on Flexible Radio, SUPELEC, Gif-sur-Yvette, France (email: merouane.debbah@supelec.fr).}
}

\markboth{Draft submitted to IEEE Transactions on Information Theory}{Shell \MakeLowercase{Draft submitted to IEEE Transactions on Information Theory}}

\maketitle

\begin{abstract}
A multi-hop relaying system is analyzed where data sent by a multi-antenna source is relayed by successive multi-antenna relays until it reaches a multi-antenna destination. Assuming correlated fading at each hop, each relay receives a faded version of the signal from the previous level, performs linear precoding and retransmits it to the next level. Using free probability theory and assuming that the noise power at relaying levels--- but not at destination--- is negligible, the closed-form expression of the asymptotic instantaneous end-to-end mutual information is derived as the number of antennas at all levels grows large. The so-obtained deterministic expression is independent from the channel realizations while depending only on channel statistics. Moreover, it also serves as the asymptotic value of the average end-to-end mutual  information. The optimal singular vectors of the precoding matrices that maximize the average mutual  information with finite number of antennas at all levels are also provided. It turns out that the optimal  precoding singular vectors are aligned to the eigenvectors of the channel correlation matrices. Thus they  can be determined using only the known channel statistics. As the optimal precoding singular vectors are  independent from the system size, they are also optimal in the asymptotic regime.
\end{abstract}


\IEEEpeerreviewmaketitle

\begin{figure*}[tbp]
\centering
\psfrag{x_0}{\LARGE{$\bx_0$}}
\psfrag{x_1}{\LARGE{$\bx_1$}}
\psfrag{x_2}{\LARGE{$\bx_2$}}
\psfrag{x_{N-2}}{\LARGE{$\bx_{N-2}$}}
\psfrag{x_{N-1}}{\LARGE{$\bx_{N-1}$}}
\psfrag{y_0}{\LARGE{$\by_0$}}
\psfrag{y_1}{\LARGE{$\by_1$}}
\psfrag{y_2}{\LARGE{$\by_2$}}
\psfrag{y_{N-2}}{\LARGE{$\by_{N-2}$}}
\psfrag{y_{N-1}}{\LARGE{$\by_{N-1}$}}
\psfrag{y_N}{\LARGE{$\by_N$}}
\psfrag{k_0}{\LARGE{$k_0$}}
\psfrag{k_1}{\LARGE{$k_1$}}
\psfrag{k_2}{\LARGE{$k_2$}}
\psfrag{k_{N-2}}{\LARGE{$k_{N-2}$}}
\psfrag{k_{N-1}}{\LARGE{$k_{N-1}$}}
\psfrag{k_N}{\LARGE{$k_N$}}
\psfrag{H_1}{\LARGE{$\bH_1$}}
\psfrag{H_2}{\LARGE{$\bH_2$}}
\psfrag{H_{N-1}}{\LARGE{$\bH_{N-1}$}}
\psfrag{H_N}{\LARGE{$\bH_N$}}
\psfrag{z}{\LARGE{$\bz$}}
\resizebox{16.5cm}{!}{\includegraphics{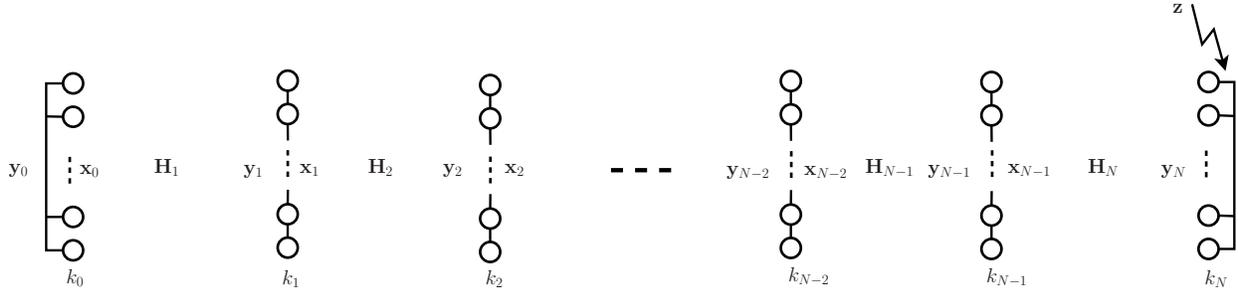}}
\caption{Multi-level Relaying System}
\label{figure1}
\end{figure*}

\section{Introduction}\label{sec:Introduction}

Relay communication systems have recently attracted much attention due to their potential to substantially improve the  signal
reception quality when the direct communication link between the source and the destination is not reliable. Due to its major
practical importance as well as its significant technical challenge, deriving the capacity - or bounds on the capacity - of various relay communication schemes is growing to an entire field of research. Of particular interest is the derivation of capacity bounds for systems in which the source, the destination, and the relays are equipped with multiple antennas.


Several works have focused on the capacity of two-hop relay networks, such as \cite{WZH,BNOP06,Morgenshtern-Bolcskei-2006,Morgenshtern-Bolcskei-Allerton2006,MB07,Li-Han-Poor-2008,VH07}.
Assuming fixed channel conditions, lower and upper bounds on the capacity of the two-hop multiple-input multiple output (MIMO) relay channel were derived in \cite{WZH}.
In the same paper, bounds on the ergodic capacity were also obtained when the communication links undergo i.i.d. Rayleigh fading.
The capacity of a MIMO two-hop relay system was studied in \cite{BNOP06} in the asymptotic case where the number of relay nodes grows large while the number of transmit and receive antennas remain constant.
The scaling behavior of the capacity in two-hop amplify-and-forward (AF) networks was analyzed in \cite{Morgenshtern-Bolcskei-2006,Morgenshtern-Bolcskei-Allerton2006,MB07} when the numbers of single-antenna sources, relays and destinations grow large.
The achievable rates of a two-hop code-division multiple-access (CDMA) decode-and-forward (DF)
relay system were derived in \cite{Cottatellucci-Chan-Fawaz-2008} when the numbers of transmit antennas and relays grow large.
In \cite{Li-Han-Poor-2008}, an ad hoc network with several source-destination pairs communicating through multiple
AF-relays was studied and an upperbound on the asymptotic capacity in the low Signal-to-Noise Ratio (SNR) regime was obtained in the case where the numbers of source, relay and destination nodes grow large. 
The scaling behavior of the capacity of a two-hop MIMO relay channel was also studied in \cite{VH07} for bi-directional transmissions.
In \cite{TH07} the optimal relay precoding matrix was derived for a two-hop relay system  with perfect knowledge of the source-relay and relay-destination channel matrices at the relay.

Following the work in \cite{Mueller-2002} on the asymptotic eigenvalue distribution of concatenated fading channels, several analysis were proposed for more general multi-hop relay networks, including \cite{Borade-Zheng-2007,Yang-Belfiore-2008,YL07,Fawaz-Zarifi-2008,Ozgur-Leveque-Tse-2007}.
In particular, considering multi-hop MIMO AF networks, the tradeoffs between rate, diversity, and network size were analyzed in \cite{Borade-Zheng-2007}, and the diversity-multiplexing tradeoff 
was derived in  \cite{Yang-Belfiore-2008}.
The asymptotic capacity of multi-hop MIMO AF relay systems was obtained in \cite{YL07} when all channel links experience i.i.d. Rayleigh fading while the number of transmit and receive antennas, as well as the number of relays at each hop grow large with the same rate.
Finally hierarchical multi-hop MIMO networks were studied in \cite{Ozgur-Leveque-Tse-2007} and the scaling laws of capacity were derived when the network density increases.


In this paper, we study an $N$-hop MIMO relay communication system wherein data transmission from $k_0$ source antennas to $k_N$ destination antennas is made possible through $N-1$ relay levels, each of which are equipped with $k_i,~i=1,\ldots,N-1$ antennas. In this transmission chain with $N+1$ levels it is assumed that the direct communication link is only viable between two adjacent levels: each relay receives a faded version of the multi-dimensional signal transmitted from the previous level and, after linear precoding, retransmits it to the next level. We consider the case where all communication links undergo Rayleigh flat fading and the fading channels at each hop (between two adjacent levels) may be correlated while the fading channels of any two different hops are independent.
We assume that the channel at each hop is block-fading and that the channel coherence-time is long enough --- with respect to codeword length --- for the system to be in the non-ergodic regime. As a consequence, the channel is a realization of a random matrix that is fixed during a coherence block, and the instantaneous end-to-end mutual information between the source and the destination is a random quantity. 

Using tools from the free probability theory and assuming that the noise power at the relay levels, but not at the destination, is negligible, we derive a closed-form expression of the asymptotic instantaneous end-to-end mutual information between the source input and the destination output as the number of antennas at all levels grows large. 
This asymptotic expression is shown to be independent from the channel realizations and to only depend on the channel statistics. Therefore, as long as the statistical properties of the channel matrices at all hops do not change, the instantaneous mutual information asymptotically converges to the same deterministic expression for any arbitrary channel realization. This property has two major consequences.
First, the mutual information in the asymptotic regime is not a random variable any more but a deterministic value representing an achievable rate. This means that when the channel is random but fixed
during the transmission and the system size is large enough, the capacity in the sense of Shannon is not zero, on the contrary to the capacity of small size systems \cite[Section 5.1]{Telatar-1995}. Second, given the stationarity of channel statistical properties, the asymptotic instantaneous mutual information obtained in the non-ergodic regime also serves as the asymptotic value of the average end-to-end mutual information between the source and the destination. Note that the latter is the same as the asymptotic ergodic end-to-end mutual information that would be obtained if the channel was an ergodic process.

We also obtain the singular vectors of the optimal precoding matrices that maximize the average mutual information of the system with a finite number of antennas at all levels. It is proven that the singular vectors of the optimal precoding matrices are also independent from the channel realizations and can be determined only using statistical knowledge of channel matrices at source and relays. We  show that the so-obtained singular vectors are also optimal in the asymptotic regime of our concern. The derived asymptotic mutual information expression and optimal precoding singular vectors set the stage for our future work on obtaining the optimal power allocation, or, equivalently, finding the optimal precoding singular values.
Finally, we apply the aforementioned results on the asymptotic mutual
information and the structure of the optimal precoding matrices to several
communications scenarios with different number of hops, and types of channel
correlation.


The rest of the paper is organized as follows. Notations and the system model are
presented in Section~\ref{sec:SysMod}. The end-to-end instantaneous mutual information in the asymptotic regime is derived in Section \ref{sec:AsymptoticMutualInfo}, while the optimal singular vectors of the precoding matrices are obtained in Section \ref{sec:OptimTxStrategy}. Theorems derived in Sections \ref{sec:AsymptoticMutualInfo} and \ref{sec:OptimTxStrategy} are applied to several MIMO communication scenarios in Section \ref{sec:ComScenario}. Numerical results are provided in Section \ref{sec:NumRes} and concluding remarks are drawn in Section \ref{sec:Conclusion}.

\section{System Model}\label{sec:SysMod}

\emph{Notation:} $\log$ denotes the logarithm in base $2$ while $\ln$ is the logarithm in base $e$.
$u(x)$ is the unit-step function defined by $u(x)=0 \mbox{ if } x < 0 \mbox{ ; } u(x)=1 \mbox{ if } x \geq 0$.
$K(m) \triangleq \int_0^{\frac{\pi}{2}} \frac{d\theta}{\sqrt{1-m \sin^2\theta}}$ is the complete elliptic integral of the first kind \cite{Abramowitz-Stegun-1964}.
Matrices and vectors are represented by boldface upper and lower cases, respectively. $\bA^T$, $\bA^\ast$, $\bA^H$ stand for
the transpose, the conjugate and the transpose conjugate of $\bA$, respectively. The trace and the determinant of $\bA$ are respectively denoted by $\tr(\bA)$ and  $\det(\bA)$.  $\lambda_{\bA}(1), \ldots, \lambda_{\bA}(n)$ represent the eigenvalues of an $n \times n$ matrix $\bA$. The operator norm of $\bA$ is defined by $\|\bA\| \triangleq \sqrt{\max_i \lambda_{\bA^H\bA}(i)}$, while the Fr{\"o}benius norm of $\bA$ is $\|\bA\|_F\triangleq\sqrt{\tr(\bA^H\bA)}$.
The $(i,j)$-th entry of matrix $\bA_k$ is written
$a^{(k)}_{ij}$. $\bI_N$ is the identity matrix of size $N$. $\E[\cdot]$ is the statistical expectation operator, $\H(X)$ the entropy of a variable $X$, and $\I(X;Y)$ the mutual information between variables $X$ and $Y$.
$F_{\bOmega}^{n}(\cdot)$ is the empirical eigenvalue distribution of an $n \times n$ square matrix $\bOmega$ with real eigenvalues, while $F_{\bOmega}(\cdot)$ and $f_{\bOmega}(\cdot)$ are respectively its asymptotic eigenvalue distribution and its eigenvalue probability density function when its size $n$ grows large.
We denote the matrix product by $\bigotimes_{i=1}^N \bA_i = \bA_1 \bA_2 \ldots \bA_N$. Note that the matrix product is not commutative, therefore the order of the index $i$ in the product is important and in particular $(\bigotimes_{i=1}^N \bA_i) ^H= \bigotimes_{i=N}^1 \bA_i^H$.

\subsection{Multi-hop MIMO relay network}

Consider Fig.~\ref{figure1} that shows a multi-hop relaying system with $k_0$ source antennas, $k_N$ destination antennas and $N-1$ relaying levels. The $i-$th  relaying level is equipped with $k_i$ antennas. We assume that the noise power is negligible at all relays while at the destination the noise power is such that
\begin{equation}\label{noise}
 \E[\bz\bz^H]=\sigma^2\bI=\frac{1}{\eta}\bI
\end{equation}
where $\bz$ is the circularly-symmetric zero-mean i.i.d. Gaussian noise vector at the destination. The simplifying noise-free relay assumption is a first step towards the future information-theoretic study of the more complex noisy relay scenario. Note that several other authors have implicitly used a similar noise-free relay assumption. For instance, in \cite{Yang-Belfiore-2008} a multi-hop AF relay network is analyzed and it is proved that the resulting colored noise at the destination can be well-approximated by white noise in the high SNR regime.
In a multi-hop MIMO relay system, it can be shown that the white-noise assumption would be equivalent to assuming negligible noise at relays, but non-negligible noise at the destination.

Throughout the paper, we assume that the correlated channel matrix at hop $i\in\{1,\ldots,N\}$ can be represented by the Kronecker model
\begin{equation}\label{channels}
 \bH_i\triangleq\bC_{r,i}^{1/2}\bTheta_i\bC_{t,i}^{1/2}
\end{equation}
where $\bC_{t,i}, \bC_{r,i}$ are respectively the transmit and receive correlation matrices, $\bTheta_i$ are zero-mean i.i.d. Gaussian matrices independent over index $i$, with variance of the $(k,l)$-th entry
\begin{equation}\label{channelvars}
 \E[|\theta^{(i)}_{kl}|^2] = \frac{a_i}{k_{i-1}}\qquad i=1,\ldots,N
\end{equation}
where $a_i=d_i^{-\beta}$ represents the pathloss attenuation with $\beta$ and $d_i$
denoting the pathloss exponent and the length of the $i$-th hop respectively.
We also assume that channels matrices $\bH_i, \; i=1,\ldots,N$ 
remain constant during a coherence block of length $L$ and vary independently from one channel coherence block to the next.

Note that the i.i.d. Rayleigh fading channel is obtained from the above Kronecker model when matrices $\bC_{t,i}$ and $\bC_{r,i}$ are set to identity.

Within one channel coherence block, the signal transmitted by the $k_0$ source antennas at time $l\in\{0,\ldots,L-1\}$ is given by the vector $ \bx_0(l) = \bP_0 \by_0(l-1)$,
where $\bP_0$ is the source precoding matrix and $\by_0$ is a zero-mean random vector with
\begin{equation}\label{y0}
 {\rm E}\{\by_0\by_0^H\}=\bI_{k_0}
\end{equation}
which implies that
\begin{equation}\label{y0K1}
{\rm E}\{\bx_0\bx_0^H\}=\bP_0 \bP_0^H.
\end{equation}

Assuming that relays work in full-duplex mode, at time $l\in\{0,\ldots,L-1\}$ the relay at level $i$ uses a precoding matrix $\bP_i$ to linearly precode its received signal $\by_i(l-1) = \bH_i \bx_{i-1}(l-1)$ and form its transmitted signal
\begin{equation}\label{xyrel}
 \bx_i(l)=\bP_i\by_i(l-1) \qquad i=0,\ldots, N-1
\end{equation}
The precoding matrices at source and relays $\bP_i,~i=0,\ldots, N-1$ are subject to the per-node long-term average power constraints
\begin{equation}\label{eq:PowConstraints}
\tr(\E[\bx_i \bx_i^H]) \leq k_i \mathcal{P}_i \qquad i=0,\ldots,N-1.
\end{equation}
The fact that $\by_i=\bH_i \bx_{i-1}$, along with the variance $\E[|\theta^{(i)}_{kl}|^2] = \frac{a_i}{k_{i-1}}$ of $\bH_i$ elements and with the power constraint $\tr(\E[\bx_{i-1} \bx_{i-1}^H]) \leq k_{i-1} \mathcal{P}_{i-1}$ on $\bx_{i-1}$, 
render the system of our concern equivalent to a system whose random channel elements $\theta^{(i)}_{kl}$ would be i.i.d. with variance $a_i$ and whose power constraint on transmitted signal $\bx_{i-1}$ would be finite and equal to $\mathcal{P}_{i-1}$. Having finite transmit power at each level, this equivalent system   shows that adding antennas, i.e. increasing the system dimension, does not imply increasing the transmit power. Nonetheless, in order to use random matrix theory tools to derive the asymptotic instantaneous mutual information in Section \ref{sec:AsymptoticMutualInfo}, the variance of random channel elements is required to be normalized by the size of the channel matrix. That is why the normalized model--- channel variance (\ref{channelvars}) and power constraint (\ref{eq:PowConstraints})--- was adopted.

It should also be noticed that choosing diagonal precoding matrices would reduce the above scheme to the simpler AF relaying strategy.

As can be observed from Fig.~\ref{figure1}, the signal received at the destination at time $l$ is given by
\begin{eqnarray}\label{sigmodl}
 \by_N(l)&=&\bH_N\bP_{N-1}\bH_{N-1}\bP_{N-2}\ldots\bH_2\bP_1\bH_1\bP_0  \by_0(l-N)  +\bz\nonumber\\
&=&\bG_N\by_0(l-N)+\bz
\end{eqnarray}
where the end-to-end equivalent channel is
\begin{eqnarray}\label{GN2}
 \bG_N&\triangleq&\bH_N\bP_{N-1}\bH_{N-1}\bP_{N-2}\ldots\bH_2\bP_1\bH_1\bP_0\nonumber\\
&=&\bC_{r,N}^{1/2}\bTheta_N\bC_{t,N}^{1/2}\bP_{N-1}\bC_{r,N-1}^{1/2}\bTheta_{N-1}\bC_{t,N-1}^{1/2}\bP_{N-2}\ldots\bC_{r,2}^{1/2}\bTheta_2\bC_{t,2}^{1/2}\bP_1\bC_{r,1}^{1/2}\bTheta_1\bC_{t,1}^{1/2}\bP_0.
\end{eqnarray}

Let us introduce the matrices
\begin{eqnarray}\label{Ms}
\bM_0&=&\bC_{t,1}^{1/2}\bP_0\nonumber\\
\bM_i&=&\bC_{t,i+1}^{1/2}\bP_i\bC_{r,i}^{1/2}\qquad i=1,\ldots,N-1\nonumber\\
\bM_N&=&\bC_{r,N}^{1/2}.
\end{eqnarray}
Then (\ref{GN2}) can be rewritten as
\begin{equation}\label{GN4}
 \bG_N =\bM_N\bTheta_N\bM_{N-1}\bTheta_{N-1}\ldots\bM_2\bTheta_2\bM_1\bTheta_1\bM_0.
\end{equation}

For the sake of clarity, the dimensions of the matrices/vectors involved in our analysis are given below.
\begin{center}
\begin{tabular}{lll}
$\bx_i: k_i\times 1$ & $\by_i: k_i\times 1$ & $\bP_i: k_i\times k_i$\\
$\bH_i: k_i\times k_{i-1}$ & $\bC_{r,i}: k_i \times k_i$ & $\bC_{t,i}: k_{i-1} \times k_{i-1}$\\
$\bTheta_i: k_i\times k_{i-1}$ & $\bM_i: k_i \times k_i$ &
\end{tabular}
\end{center}

In the sequel, we assume that the channel coherence time is large enough to consider the non-ergodic case and consequently, time index $l$ can be dropped. Finally, we define three channel-knowledge assumptions:
\begin{itemize}
  \item Assumption $\bAs$, local statistical knowledge at source: the source has only statistical channel state information (CSI) of its forward channel $\bH_1$, i.e. the source knows the transmit correlation matrix $\bC_{t,1}$.
  \item Assumption $\bAr$, local statistical knowledge at relay: at the $i^{th}$ relaying level, $i\in\{1,\ldots,N-1\}$,
   only statistical CSI of the backward channel $\bH_i$ and forward channel $\bH_{i+1}$ are available, i.e. relay $i$ knows the receive correlation matrix $\bC_{r,i}$ and the transmit correlation matrix $\bC_{t,i+1}$.
  \item Assumption $\bAd$, end-to-end perfect knowledge at destination: the destination perfectly knows the end-to-end equivalent channel $\bG_N$.
\end{itemize}
Throughout the paper, assumption $\bAd$ is always made. Assumption $\bAd$ is the single assumption on channel-knowledge necessary to derive the asymptotic mutual information in Section \ref{sec:AsymptoticMutualInfo}, while the two extra assumptions $\bAs$ and $\bAr$ are also necessary in Section \ref{sec:OptimTxStrategy} to obtain the singular vectors of the optimal precoding matrices.

\subsection{Mutual Information}

Consider the channel realization $\bG_N$ in one channel coherence block. Under Assumption $\bAd$, the instantaneous end-to-end mutual information between channel input $\by_0$ and channel output $(\by_N,\bG_N)$ in this channel coherence block is \cite{Telatar-1995}
\begin{equation}\label{eq:instMutInf}
\begin{split}
\I(y_0;y_N | G_N=\bG_N)
&= \H(y_N| G_N=\bG_N)-\underbrace{\H(y_N | y_0,G_N=\bG_N)}_{\H(z)}\\
&= \H(y_N | G_N = \bG_N) - \H(z)
\end{split}
\end{equation}
The entropy of the noise vector is known to be $\H(z)=\log\det(\frac{\pi e}{\eta} \bI_{k_N})$.
Besides, $\by_0$ is zero-mean with variance $\E[\by_0\by_0^H]=\bI_{k_0}$, thus given $\bG_N$, the received signal $\by_N$ is zero-mean with variance {$\bG_N \bG_N^H + \frac{1}{\eta} \bI_{k_N}$}. By \cite[Lemma 2]{Telatar-1995}, we have the inequality $\H(y_N| G_N = \bG_N)\leq \log \det(\pi e \bG_N \bG_N^H + \frac{\pi e}{\eta} \bI_{k_N})$, and the entropy is maximized when the latter inequality holds with equality. This occurs if $\by_N$ is circularly-symmetric complex Gaussian, which is the case when $\by_0$ is circularly-symmetric complex Gaussian. Therefore throughout the rest of the paper we consider $\by_0$ to be zero-mean a circularly-symmetric complex Gaussian vector. As such, the instantaneous mutual information (\ref{eq:instMutInf}) can be rewritten as
\begin{equation}\label{eq:InstMutualInfo}
\begin{split}
\I(y_0;y_N | G_N=\bG_N)
&= \log\det(\bI_{k_N} + \eta \bG_N \bG_N^H ).
\end{split}
\end{equation}

Under Assumption $\bAd$, the average end-to-end mutual information between channel input $\by_0$ and channel output $(\by_N,\bG_N)$ is
\begin{equation}\label{eq:mutInf}
\begin{split}
\I(y_0;(y_N,G_N))
& =\I(y_0;y_N | G_N)+ \underbrace{\I(y_0;G_N)}_{0} \\
&= \I(y_0;y_N | G_N) \\
&= \E_{G_N} [\I(y_0;y_N | G_N = \bG_N)]\\
& = \E_{G_N} [\log\det(\bI_{k_N} + \eta \bG_N \bG_N^H )] .
\end{split}
\end{equation}
To optimize the system, we are left with finding the precoders $\bP_i$ that maximize the end-to-end mutual information (\ref{eq:mutInf}) subject to power constraints (\ref{eq:PowConstraints}). In other words, we need to find the maximum average end-to-end mutual information
\begin{equation}\label{eq:Capa}
\begin{split}
 \bbC & \triangleq \max_{\{\bP_i / \tr(\E[\bx_i \bx_i^H]) \leq k_i \mathcal{P}_i \}_{i\in\{0,\ldots,N-1\}}}
 \E_{G_N}\left[\log \det(\bI_{k_N}+ \eta \; \bG_N\bG_N^H)\right]
\end{split}
\end{equation}
Note that the non-ergodic regime is considered, therefore (\ref{eq:mutInf}) represents only an average mutual information over channel realizations, and the solution to (\ref{eq:Capa}) does not necessarily represent the channel capacity in the Shannon sense when the system size is small.

\section{Asymptotic Mutual Information}\label{sec:AsymptoticMutualInfo}

In this section, we consider the instantaneous mutual information per source antenna between the source and the destination
\begin{equation}
 \bbI\triangleq\frac{1}{k_0}\log\det(\bI_{k_N}+\eta\bG_N\bG_N^H)
\end{equation}
and derive its asymptotic value as the number of antennas $k_0, k_1,\ldots, k_N$ grow large. The following theorem holds.

\begin{theorem}\label{th:Iasymptotic}
For the system described in section \ref{sec:SysMod}, assume that
\begin{itemize}
  \item channel knowledge assumption $\bAd$ holds;
  \item $k_0, k_1,\ldots, k_N \rightarrow \infty$ while $ \frac{k_i}{k_N}\rightarrow\rho_i$ for $i=0,\ldots, N$;
  \item for $i = 0,\ldots, N$, as $k_i\rightarrow \infty$, $\bM_i^H\bM_i$ has a limit eigenvalue distribution with a compact support. 
\end{itemize}
Then the instantaneous mutual information per source antenna $\bbI$ converges almost surely to
\begin{equation}\label{eq:hh1}
\bbI_{\infty}
=\frac{1}{\rho_0}\sum_{i=0}^N\rho_i\E\left[\log\left(1+\eta \frac{a_{i+1}}{\rho_i} h_i^N\Lambda_i\right)\right]-N\frac{\log e}{\rho_0}\eta\prod_{i=0}^N h_i
\end{equation}
where $a_{N+1}=1$ by convention, $h_0,h_1,\ldots,h_N$ are the solutions of the system of $N+1$ equations
\begin{equation}\label{eq:hh2}
\prod_{j=0}^N h_j=\rho_i\E\left[\frac{h_i^N\Lambda_i}{ \frac{\rho_i}{a_{i+1}} + \eta h_i^N\Lambda_i}\right] \qquad i=0,\ldots,N
\end{equation}
and the expectation $\E[\cdot]$ in (\ref{eq:hh1}) and (\ref{eq:hh2}) is over $\Lambda_i$ whose distribution is given by the asymptotic eigenvalue distribution $F_{\bM_i^H\bM_i}(\lambda)$ of $\bM_i^H\bM_i$.
\end{theorem}

\bigskip

The detailed proof of \emph{Theorem \ref{th:Iasymptotic}} is presented in Appendix \ref{ap:ProofThAsymptoticMutInfo}.

We would like to stress that (\ref{eq:hh1}) holds for any arbitrary set of precoding matrices $\bP_i,~i=0,\ldots,N-1$, if $\bM_i^H\bM_i$ has a compactly supported asymptotic eigenvalue distribution when the system dimensions grow large.
We would like to point out that the power constraints on signals transmitted by the source or relays are not sufficient to guarantee the boundedness of the eigenvalues of $\bM_i^H\bM_i$. In fact, as (\ref{eq:PowConsPi}) in Appendix \ref{ap:ProofTxDirec} shows, in the asymptotic regime the power constraints impose upper-bounds on the product of first-order moment of the eigenvalues of matrices $\bP_i \bC_{r,i} \bP_i^H$ and $\bM_k^H \bM_k$--- indeed $ \lim_{k_i \rightarrow \infty} \frac{1}{k_i} \tr(\bP_i \bC_{r,i} \bP_i^H) = \E[\lambda_{\bP_i \bC_{r,i} \bP_i^H}]$ and $ \lim_{k_k \rightarrow \infty} \frac{1}{k_k} \tr(\bC_{t,k+1} \bP_k \bC_{r,k} \bP_k^H) = \E[\Lambda_k]$. Unfortunately, these upper-bounds do not prevent the eigenvalue distribution of $\bM_i^H\bM_i$ from having an unbounded support.
Thus, the assumption that matrices $\bM_i^H\bM_i$ have a compactly supported asymptotic eigenvalue distribution is a priori not an intrinsic property of the system model, and it was necessary to make that assumption in order to use \emph{Lemma \ref{lem:Hiai}} to prove \emph{Theorem \ref{th:Iasymptotic}}.

Given a set of precoding matrices, it can be observed from (\ref{eq:hh1}) and (\ref{eq:hh2}) that the asymptotic expression is a deterministic value that depends only on channel statistics and not on a particular channel realization. In other words, for a given set of precoding matrices, as long as the statistical properties of the channel matrices do not change, the instantaneous mutual information always converges to the same deterministic achievable rate, regardless of the channel realization. Thus, as the numbers of antennas at all levels grow large, the instantaneous mutual information is not a random variable anymore and the precoding matrices maximizing the asymptotic instantaneous mutual information can be found based only on knowledge of the channel statistics, without requiring any information regarding the instantaneous channel realizations.
This further means that when the channel is random but fixed during the transmission and the system size grows large enough, the Shannon capacity is not zero any more, on the contrary to the capacity of small-size systems \cite[Section 5.1]{Telatar-1995}.

Moreover, given the stationarity of channel statistical properties, the instantaneous mutual information converges to the same deterministic expression for any arbitrary channel realization. Therefore, the asymptotic instantaneous mutual information (\ref{eq:hh1}) obtained in the non-ergodic regime also represents the asymptotic value of the average mutual information, whose expression is the same as the asymptotic ergodic end-to-end mutual information that would be obtained if the channel was an ergodic process.

It should also be mentioned that, according to the experimental results illustrated in Section \ref{sec:NumRes}, the system under consideration behaves like in the asymptotic regime even when it is equipped with a reasonable finite number of antennas at each level. Therefore, (\ref{eq:hh1}) can also be efficiently used to evaluate the instantaneous mutual information of a finite-size system.

\section{Optimal Transmission Strategy at Source and Relays}\label{sec:OptimTxStrategy}

In previous section, the asymptotic instantaneous mutual information (\ref{eq:hh1}), (\ref{eq:hh2}) was derived considering arbitrary precoding matrices $\bP_i, i\in\{0,\ldots,N-1\}$.
In this section, we analyze the optimal linear precoding strategies $\bP_i, i\in\{0,\ldots,N-1\}$ at source and relays that allow to maximize the average mutual information.
We characterize the optimal transmit directions determined by the singular vectors of the precoding matrices at source and relays, for a system with finite $k_0, k_1,\ldots, k_N$. It turns out that those transmit direction are also the ones that maximize the asymptotic average mutual information. As explained in Section \ref{sec:AsymptoticMutualInfo}, in the asymptotic regime, the average mutual information and the instantaneous mutual information have the same asymptotic value, therefore the singular vectors of the precoding matrices maximizing the asymptotic average mutual information are also optimal for the asymptotic instantaneous mutual information (\ref{eq:hh1}).

In future work, using the results on the optimal directions of transmission (singular vectors of $\bP_i$) and the asymptotic mutual information (\ref{eq:hh1})--(\ref{eq:hh2}), we intend to derive the optimal power allocation (singular values of $\bP_i$) that maximize the asymptotic instantaneous/average mutual information (\ref{eq:hh1}) using only statistical knowledge of the channel at transmitting nodes.

The main result of this section is given by the following theorem:

\begin{theorem}\label{th:txDirections}
Consider the system described in Section \ref{sec:SysMod}.
For $i\in\{1,\ldots,N\}$ let $\bC_{t,i}=\bU_{t,i}\bLambda_{t,i}\bU_{t,i}^H$ and $\bC_{r,i}=\bU_{r,i}\bLambda_{r,i}\bU_{r,i}^H$ be the eigenvalue decompositions of the correlation matrices $\bC_{t,i}$ and $\bC_{r,i}$, where $\bU_{t,i}$ and $\bU_{r,i}$ are unitary and $\bLambda_{t,i}$ and $\bLambda_{r,i}$ are diagonal, with their respective eigenvalues ordered in decreasing order.
Then, under channel-knowledge assumptions $\bAs$, $\bAr$ and $\bAd$, the optimal linear precoding matrices that maximize the average mutual information under power constraints (\ref{eq:PowConstraints}) can be written as
\begin{equation}\label{eq:PrecVec}
\begin{split}
\bP_0 & =\bU_{t,1}\bLambda_{P_0} \\
\bP_i & =\bU_{t,i+1}\bLambda_{P_i}\bU_{r,i}^H \mbox{ , for } i\in\{1,\ldots,N-1\}
\end{split}
\end{equation}
where $\bLambda_{P_i}$ are diagonal matrices with non-negative real diagonal elements. 
Moreover, the singular vectors of the precoding matrices (\ref{eq:PrecVec}) are also the ones that maximize the asymptotic average mutual information.
Since the asymptotic average mutual information has the same value as the asymptotic instantaneous mutual information, the singular vectors of the precoding matrices (\ref{eq:PrecVec}) are also optimal for the asymptotic instantaneous mutual information.
\end{theorem}

\bigskip

For the proof of \emph{Theorem \ref{th:txDirections}}, the reader is referred to Appendix \ref{ap:ProofTxDirec}.

\emph{Theorem \ref{th:txDirections}} indicates that to maximize the average mutual information
\begin{itemize}
  \item the source should align the eigenvectors of the transmit covariance matrix $\bQ =\bP_0 \bP_0^H$ to the eigenvectors of the transmit correlation matrix $\bC_{t,1}$ of the first-hop channel $\bH_1$. This alignment requires only local statistical channel knowledge $\bAs$.
      Note that similar results were previously obtained for both single-user \cite{Jafar-Goldsmith-2004} and multi-user \cite{Soysal-Ulukus-2007} single-hop (without relays) MIMO system with covariance knowledge at the source.
  \item   relay $i$ should align the right singular vectors of its precoding matrix $\bP_i$ to the eigenvectors of the receive correlation matrix $\bC_{r,i}$, and the left singular vectors of $\bP_i$ to the eigenvectors of the transmit correlation matrix $\bC_{t,i+1}$. These alignments require only local statistical knowledge $\bAr$.
\end{itemize}
Moreover, it follows from \emph{Theorem \ref{th:txDirections}} that the optimization of $\bP_i$ can be divided into two decoupled problems: optimizing the transmit directions---singular vectors--- on one hand, and optimizing the transmit powers---singular values--- on the other hand.

We would like to draw the reader's attention to the fact that the proof of this theorem does not rely on the expression of the asymptotic mutual information given in (\ref{eq:hh1}).
In fact, \emph{Theorem \ref{th:txDirections}} is first proved in the non-asymptotic regime for an arbitrary set of $\{k_i\}_{i\in\{0,\ldots,N\}}$.
As such, regardless of the system size, the singular vectors of the precoding matrices should always be aligned to the eigenvectors of the channel correlation matrices to maximize the average mutual information. In particular,  the singular vectors of the precoding matrices that maximize the asymptotic average mutual information are also aligned to the eigenvectors of channel correlation matrices as in (\ref{eq:PrecVec}). As explained in Section \ref{sec:AsymptoticMutualInfo},  the instantaneous and the average mutual informations have the same value in the asymptotic regime. Therefore, the singular vectors given in (\ref{eq:PrecVec}) are also those that maximize the asymptotic instantaneous mutual information.

\section{Application to MIMO Communication Scenarios}\label{sec:ComScenario}

In this section,  \emph{Theorem}~\ref{th:Iasymptotic} and  \emph{Theorem}~\ref{th:txDirections} are applied to four different communication scenarios. In the first two scenarios, the special case of non-relay assisted MIMO (N=1) without path-loss ($a_1=1$) is considered, and we show how (\ref{eq:hh1}) boils down to known results for the MIMO channel with or without correlation. In the third and fourth scenarios, a multi-hop MIMO system is considered and the asymptotic mutual information is developed in the uncorrelated and exponential correlation cases respectively.

\subsection{Uncorrelated single-hop MIMO with statistical CSI at source}\label{sec:ComScen1}

Consider a simple single-hop uncorrelated MIMO system with the same number of antennas at source and destination i.e. $\rho_0 = \rho_1 = 1 $, and an i.i.d. Rayleigh fading channel i.e. $\bC_{t,1} = \bC_{r,1} = \bI$. Assuming equal power allocation at source antennas, the source precoder is $\bP_0 = \sqrt{\mathcal{P}_0} \bI$. As $\bM_0 = \bC_{t,1}^{1/2}\bP_0 = \sqrt{\mathcal{P}_0} \bI$ and $\bM_1= \bC_{r,1}^{1/2}  = \bI$, we have that
\begin{equation}\label{ex1}
\begin{split}
dF_{\bM_0^H\bM_0}(\lambda) &= \delta \left(\lambda - \mathcal{P}_0\right) d\lambda\\
dF_{\bM_1^H\bM_1}(\lambda) &= \delta (\lambda - 1) d\lambda.
\end{split}
\end{equation}

Using the distributions in (\ref{ex1}) to compute the expectations in (\ref{eq:hh1}) yields
\begin{equation}\label{ex2}
\begin{split}
\bbI_{\infty} &=\frac{1}{\rho_0}\sum_{i=0}^N\rho_i\E\left[\log\left( 1 + \frac{\eta}{\rho_i} h_i^N \Lambda_i \right)\right] -N\frac{\log e}{\rho_0}\eta\prod_{i=0}^N h_i\\
&= \log\left(1 + \eta h_0 \mathcal{P}_0 \right) + \log (1 + \eta h_1) - \log e \:\: \eta \: h_0 \: h_1
\end{split}
\end{equation}
where, according to (\ref{eq:hh2}), $h_0$ and $h_1$ are the solutions of the system of two 
equations
\begin{equation}\label{ex3}
\begin{split}
  h_0 &= \frac{1}{1+\eta h_1} \\
  h_1 &= \frac{{\mathcal{P}_0}}{1+\eta h_0 {\mathcal{P}_0}}
\end{split}
\end{equation}
that are given by
\begin{equation}\label{ex4}
\begin{split}
  h_0 &= \frac{2}{1+\sqrt{1+4\eta {\mathcal{P}_0}}} \\
  h_1 &= \frac{-1+\sqrt{1+4\eta \mathcal{P}_0}}{2\eta}.
\end{split}
\end{equation}
Using (\ref{ex4}) in (\ref{ex2}), we obtain
\begin{equation}\label{eq:MIMOiidCapa}
\bbI_{\infty} = 2 \log \left( \frac{1 + \sqrt{1+4\eta \mathcal{P}_0 }}{2} \right) - \frac{\log e}{4 \eta \mathcal{P}_0} \left(\sqrt{1+4\eta \mathcal{P}_0 } - 1 \right)^2.
\end{equation}

It can be observed that the deterministic expression (\ref{eq:MIMOiidCapa}) depends only on the system characteristics and is independent from the channel realizations.
Moreover, equal power allocation is known to be the capacity-achieving power allocation for a MIMO i.i.d. Rayleigh channel with statistical CSI at source \cite[Section 3.3.2]{Tulino-Verdu-2004},\cite{Telatar-1995}.
As such, the asymptotic mutual information (\ref{eq:MIMOiidCapa}) also represents the asymptotic capacity of the system.
We should also mention that (\ref{eq:MIMOiidCapa}) is similar to the expression of the asymptotic capacity per dimension previously derived in \cite[Section 3.3.2]{Tulino-Verdu-2004} for the MIMO Rayleigh channel with equal number of transmit and receive antennas and statistical CSI at the transmitter.

\subsection{Correlated single-hop MIMO with statistical CSI at source}\label{sec:ComScen2}

In this example, we consider the more general case of correlated MIMO channel with separable correlation:
$\bH_1=\bC_{r,1}^{1/2}\bTheta_1\bC_{t,1}^{1/2}$.
Let us denote the eigenvalue decomposition of $\bC_{t,1}$ as
\begin{equation}\label{xs0}
\bC_{t,1}=\bU_{t,1} \bLambda_{t,1} \bU_{t,1}^H
\end{equation}
where $\bLambda_{t,1}$ is a diagonal matrix whose diagonal entries are the eigenvalues of $\bC_{t,1}$ in the non-increasing order and the unitary matrix $\bU_{t,1}$ contains the corresponding eigenvectors. Defining the transmit covariance matrix
\begin{equation}\label{xs1}
\bQ\triangleq \E\left[\bx_0\bx_0^H\right]=\bP_0\bP_0^H,
\end{equation}
it has been shown \cite{Jafar-Goldsmith-2004} that the capacity-achieving matrix
$\bQ^\star$ is given by
\begin{equation}\label{xs2}
\bQ^\star=\bU_{t,1}\bLambda_{\bQ^\star}\bU_{t,1}^H
\end{equation}
where $\bLambda_{\bQ^\star}$ is a diagonal matrix containing the capacity-achieving power allocation.
Using \emph{Theorem~\ref{th:Iasymptotic}} along with (\ref{xs0}) and (\ref{xs2}), it can be readily shown that the asymptotic capacity per dimension is equal to
\begin{equation}\label{k1}
C = \E[\log (1+ \frac{\eta}{\rho_0} \Lambda_0 h_0)]+\frac{1}{\rho_0}\E[\log (1+ \eta \Lambda_1 h_1)]  - \frac{\log e }{\rho_0} \eta \: h_0 h_1
\end{equation}
where $h_0$ and $h_1$ are the solutions of the system
\begin{equation}\label{xs3}
\begin{split}
h_0  &=  \E \left[\frac{\Lambda_1}{1+ \eta \Lambda_1 h_1}\right]\\
h_1  &= \E \left[\frac{\Lambda_0}{1+ \frac{\eta}{\rho_0} \Lambda_0 h_0}\right]
\end{split}
\end{equation}
and the expectations are over $\Lambda_0$ and $\Lambda_1$ whose distributions are given by the asymptotic eigenvalue distributions of $\bLambda_{t,1} \bLambda_{\bQ^\star}$ and $\bC_r$, respectively.
It should be mentioned that an equivalent expression\footnote{The small differences between (\ref{k1}) and the capacity expression in \cite[Theorem~3.7]{Tulino-Verdu-2004} are due to different normalization assumptions in \cite{Tulino-Verdu-2004}. In particular (\ref{k1}) is the mutual information per source antenna while the expression in \cite{Tulino-Verdu-2004} is the capacity per receive antenna.
The equivalence between  \cite[Theorem~3.7]{Tulino-Verdu-2004} and (\ref{k1}) is obtained according to the following notation equivalence (\{\cite{Tulino-Verdu-2004}-notation\} $\sim$ \{(\ref{k1})-notation\}):
\begin{equation}
\begin{split}
C & \sim \rho_0 \bbI_{\infty} \qquad \beta  \sim \rho_0 \qquad \SNR \sim  \mathcal{P}_0 \eta  \qquad \Gamma  \sim  \frac{h_0}{\rho_0} \qquad \Upsilon  \sim \frac{h_1}{\mathcal{P}_0}\\
\Lambda_R & \sim \Lambda_1  \mbox{ , both with distribution given by the eigenvalue distribution of } \bC_r  \\
\Lambda & \sim \frac{\Lambda_0}{\mathcal{P}_0} \mbox{ , both with distribution given by the eigenvalue distribution of } \bLambda_{t,1}\bLambda_{\bQ^\star}/\mathcal{P}_0
\end{split}
\end{equation}
} was obtained in \cite[Theorem~3.7]{Tulino-Verdu-2004} for the capacity of the correlated MIMO channel with statistical CSI at transmitter.

\subsection{Uncorrelated multi-hop MIMO with statistical CSI at source and relays}\label{sec:ComScen3}

In this example, we consider an uncorrelated multi-hop MIMO system, i.e. all correlation matrices are equal to identity. Then by \emph{Theorem \ref{th:txDirections}} the optimal precoding matrices should be diagonal. Assuming equal power allocation at source and relays, the precoding matrices are of the form  $\bP_i=\alpha_i \bI_{k_i}$, where $\alpha_i$ is real positive and chosen to respect the power constraints.

Using the power constraint expression (\ref{eq:PowConsPi}) in Appendix \ref{ap:ProofTxDirec}, it can be shown by induction on $i$ that the coefficients $\alpha_i$ in the uncorrelated case are given by
\begin{equation}\label{eq:UncorrAlphai}
\begin{split}
\alpha_0 & = \sqrt{\mathcal{P}_0}\\
\alpha_i & = \sqrt{\frac{\mathcal{P}_i}{a_i \mathcal{P}_{i-1}}} \qquad \forall i \in \{1,\ldots,N-1\}\\
\alpha_N & = 1.
\end{split}
\end{equation}

Then the asymptotic mutual information for the uncorrelated multi-hop MIMO system with equal power allocation is given by
\begin{equation}
\bbI_{\infty}
=\sum_{i=0}^N \frac{\rho_i}{\rho_0} \log\left(1+ \frac{\eta h_i^N a_{i+1}\alpha_i^2}{\rho_i} \right)-N\frac{\log e}{\rho_0}\eta\prod_{i=0}^N h_i
\end{equation}
where $h_0,h_1,\ldots,h_N$ are the solutions of the system of $N+1$ multivariate polynomial equations
\begin{equation}
\prod_{j=0}^N h_j= \frac{h_i^N\alpha_i^2 a_{i+1}}{ 1 + \frac{\eta h_i^N a_{i+1} \alpha_i^2}{\rho_i}}  \qquad i=0,\ldots,N.
\end{equation}
Note that the asymptotic mutual information is a deterministic value depending only on a few system characteristics: signal power $\mathcal{P}_i$, noise power $1/\eta$, pathloss $a_i$, number of hops $N$ and ratio of the number of antennas $\rho_i$.

\subsection{Exponentially correlated multi-hop MIMO with statistical CSI at source and relays}\label{sec:ComScen4}

In this example, the asymptotic mutual information (\ref{eq:hh1}) is developed in the case of exponential correlation matrices and precoding matrices with optimal singular vectors.

\noindent\textbf{Optimal precoding directions:}
For $i\in\{1,\ldots,N\}$, the eigenvalue decompositions of channel correlation matrices $\bC_{t,i}$ and $\bC_{r,i}$ can be written as
\begin{equation}
\begin{split}
\bC_{t,i} & =\bU_{t,i}\bLambda_{t,i}\bU_{t,i}^H \\
\bC_{r,i} & =\bU_{r,i}\bLambda_{r,i}\bU_{r,i}^H
\end{split}
\end{equation}
where $\bU_{t,i}$ and $\bU_{r,i}$ are unitary, and $\bLambda_{t,i}$ and $\bLambda_{r,i}$ are diagonal with their respective eigenvalues ordered in decreasing order.
Following \emph{Theorem \ref{th:txDirections}}, we consider precoding matrices of the form $\bP_i=\bU_{t,i+1}\bLambda_{P_i}\bU_{r,i}^H$,
i.e. the singular vectors of $\bP_i$ are optimally aligned to the eigenvectors of channel correlation matrices.

Consequently, we can rewrite matrices $\bM_i^H\bM_i$ (\ref{Ms}) as
\begin{equation}
\begin{split}
\bM_0^H \bM_0&=\bU_{t,1}^H\bLambda_{P_0}^2\bLambda_{t,1}\bU_{t,1} \\
\bM_i^H \bM_i &=\bU_{r,i}^H\bLambda_{r,i}\bLambda_{P_i}^2 \bLambda_{t,i+1}\bU_{r,i}\qquad i=1,\ldots,N-1 \\
\bM_N^H \bM_N &=\bU_{r,N}^H \bLambda_{r,N} \bU_{r,N}.
\end{split}
\end{equation}
Thus, the eigenvalues of matrices $\bM_i^H\bM_i$ are contained in the following diagonal matrices
\begin{equation}\label{eq:Lambdai}
\begin{split}
\bLambda_0 &=\bLambda_{P_0}^2\bLambda_{t,1}\\
\bLambda_i &=\bLambda_{r,i}\bLambda_{P_i}^2\bLambda_{t,i+1} \qquad i=1,\ldots,N-1 \\
\bLambda_N &= \bLambda_{r,N}.
\end{split}
\end{equation}

The asymptotic mutual information, given by (\ref{eq:hh1}) and (\ref{eq:hh2}), involves expectations of functions of $\Lambda_i$ whose distribution is given by the asymptotic eigenvalue distribution $F_{\bM_i^H\bM_i}(\lambda)$ of $\bM_i^H\bM_i$. Equation (\ref{eq:Lambdai}) shows that a function $g_1(\Lambda_i)$ can be written as a function $g_2(\Lambda_{P_i}^2, \: \Lambda_{r,i}, \: \Lambda_{t,i+1})$, where the variables $\Lambda_{P_i}^2$,  $\Lambda_{r,i}$, and $\Lambda_{t,i+1}$ are respectively characterized by the asymptotic eigenvalue distributions $F_{\bP_i^H \bP_i}(\lambda)$,  $F_{\bC_{r,i}}(\lambda)$, and  $F_{\bC_{t,i+1}}(\lambda)$ of matrices $\bP_i^H \bP_i$ , $\:\bC_{r,i}$ and $\bC_{t,i+1}$ respectively.
Therefore expectations in (\ref{eq:hh1}) and (\ref{eq:hh2}) can be computed using the asymptotic joint distribution of $(\Lambda_{P_i}^2, \: \Lambda_{r,i}, \: \Lambda_{t,i+1})$ instead of the distribution $F_{\bM_i^H\bM_i}(\lambda)$.
To simplify notations, we rename the variables as follows
\begin{equation}
X =\Lambda_{P_i}^2 \qquad Y =\Lambda_{r,i} \qquad  Z =\Lambda_{t,i+1}.
\end{equation}

Then, the expectation of a function $g_1(\Lambda_i)$ can be written
\begin{equation}\label{eq:Eg1}
\begin{split}
\E [g_1(\Lambda_i)]
& = \E [g_2(X, Y, Z)]= \int_z \int_y \int_x  { g_2(x,y,z) f_{X,Y,Z}(x,y,z) \: dx \: dy \: dz}\\
& = \int_z \int_y \int_x  { g_2(x,y,z) f_{X|Y,Z}(x|y,z) \:\: f_{Y|Z}(y|z) \:\: f_{Z}(z) \: dx \: dy \: dz}.
\end{split}
\end{equation}

\vspace*{2mm}

\noindent \textbf{Exponential Correlation Model:}
So far, general correlation matrices were considered. We now introduce the exponential correlation model and further develop (\ref{eq:Eg1}) for the distributions $f_{Y|Z}(y|z)$ and $f_{Z}(z)$ resulting from that particular correlation model.

We assume that Level $i$ is equipped with a uniform linear array (ULA) of length $L_i$, characterized by its antenna spacing $l_i=L_{i}/k_{i}$ and its characteristic distances $\Delta_{t,i}$ and $\Delta_{r,i}$ proportional to transmit and receive spatial coherences respectively.
Then the receive and transmit correlation matrices at Level $i$ can respectively be modeled by the following Hermitian Wiener-class\footnote{A sequence of $n \times n$ Toeplitz Matrices $\bT_n=[t_{k-j}]_{n\times n}$ is said to be in the Wiener class \cite[Section 4.4]{Gray-2006} if the sequence $\{t_k\}$ of first-column and first-row elements is absolutely summable, i.e. $\lim_{n\rightarrow +\infty} \sum_{k=-n}^{n} |t_k| < + \infty$.\\
If $|r_{r,i}|<1$, then $\lim_{k_i \rightarrow +\infty} (\sum_{k=0}^{k_i-1} r_{r,i}^k + \sum_{k=-k_{i-1}}^{-1} r_{r,i}^{-k}) = \frac{1}{1-r_{r,i}}+ \frac{1/r_{r,i}}{1-1/r_{r,i}} < \infty$, and consequently $\bC_{r,i}$ is in the Wiener class. $\bC_{t,i}$ is obviously also in the Wiener class if $|r_{t,i}|<1$.} Toeplitz matrices \cite{Loyka-2001,Martin-Ottersten-2004,Oestges-Clerckx-Debbah-2007}:
\begin{equation}\label{eq:ToeplitzCorr}
\bC_{r,i}=\left[
            \begin{array}{*{6}{c}}
            1                   & r_{r,i} & r_{r,i}^2    & \ldots & r_{r,i}^{k_{i}-1} \\
            r_{r,i}             & 1       & \ddots       & \ddots & \vdots              \\
            r_{r,i}^2           & \ddots  & \ddots       & \ddots & r_{r,i}^2           \\
            \vdots              & \ddots  & \ddots       & 1      & r_{r,i}             \\
            r_{r,i}^{k_{i}-1} & \ldots  & r_{r,i}^2    & r_{r,i}& 1
            \end{array}
          \right]_{k_{i} \times k_{i}}
\!\!\!\!\!\! \mbox{ and } \:\:\:\:\:\:\:\:\:
\bC_{t,i+1}=\left[
            \begin{array}{*{6}{c}}
            1                   & r_{t,i+1} & r_{t,i+1}^2  & \ldots   & r_{t,i+1}^{k_{i}-1} \\
            r_{t,i+1}           & 1         & \ddots       & \ddots   & \vdots              \\
            r_{t,i+1}^2         & \ddots    & \ddots       & \ddots   & r_{t,i+1}^2           \\
            \vdots              & \ddots    & \ddots       & 1        & r_{t,i+1}             \\
            r_{t,i+1}^{k_{i}-1} & \ldots    & r_{t,i+1}^2  & r_{t,i+1}& 1
            \end{array}
          \right]_{k_{i} \times k_{i}}
\end{equation}
where the antenna correlation at receive (resp. transmit) side $r_{r,i}=e^{-\frac{l_{i}}{\Delta_{r,i}}} \in [0,1)$  (resp. $r_{t,i+1}=e^{-\frac{l_{i}}{\Delta_{t,i}}} \in [0,1)$) is an exponential function of antenna spacing  $l_i$ and characteristic distance $\Delta_{r,i}$ (resp. $\Delta_{t,i}$ ) at relaying Level $i$.

As $k_i$ grows large, the sequence of Toeplitz matrices $\bC_{r,i}$ of size $k_i \times k_i$ is fully characterized by the continuous real function $f_{r,i}$, defined for $\lambda \in [0,2\pi)$ by \cite[Section 4.1]{Gray-2006}
\begin{equation}\label{eq:fri}
\begin{split}
f_{r,i}(\lambda)
& = \lim_{k_i \rightarrow +\infty} \left( \sum_{k=0}^{k_i-1} r_{r,i}^k e^{jk\lambda} + \sum_{k=-(k_i-1)}^{-1} r_{r,i}^{-k} e^{jk\lambda} \right)\\
& = \frac{1}{1-r_{r,i}e^{j\lambda}} + \frac{r_{r,i} e^{-j\lambda}}{1-r_{r,i}e^{-j\lambda}}\\
& = \frac{1-r_{r,i}^2}{|1-r_{r,i}e^{j\lambda}|^2}. 
\end{split}
\end{equation}
We also denote the essential infimum and supremum of $f_{r,i}$ by $m_{f_{r,i}}$ and $M_{f_{r,i}}$ respectively \cite[Section 4.1]{Gray-2006}. In a similar way, we can define the continuous real function $f_{t,i+1}$ characterizing the sequence of Toeplitz matrices $\bC_{t,i+1}$ by replacing $r_{r,i}$ in (\ref{eq:fri}) by $r_{t,i+1}$, and we denote by $m_{f_{t,i+1}}$ and $M_{f_{t,i+1}}$ its essential infimum and supremum respectively.

By Szeg{\"o} Theorem \cite[Theorem 9]{Gray-2006}, recalled hereafter in \emph{Lemma \ref{lem:Szego}}, for any real function $g(\cdot)$ (resp. $h(\cdot)$) continuous on $[m_{f_{r,i}},M_{f_{r,i}}]$ (resp. $[m_{f_{t,i+1}},M_{f_{t,i+1}}]$), we have
\begin{equation}\label{eq:Szego}
\begin{split}
\int_y g(y) f_Y(y) \: dy \triangleq \lim_{k_i \rightarrow +\infty } \frac{1}{k_i} \sum_{k=1}^{k_i} g\left(\lambda_{\bC_{r,i}}(k)\right) = \frac{1}{2\pi} \int_0^{2\pi} g\left(f_{r,i}(\lambda)\right) \; d\lambda \\
\int_z h(z) f_Z(z) \: dz \triangleq \lim_{k_i \rightarrow +\infty } \frac{1}{k_i} \sum_{k=1}^{k_i} h\left(\lambda_{\bC_{t,i+1}}(k)\right) = \frac{1}{2\pi} \int_0^{2\pi} h\left(f_{t,i+1}(\nu)\right) \; d\nu.
\end{split}
\end{equation}

Assuming that variables $Y =\Lambda_{r,i}$ and $Z =\Lambda_{t,i+1}$ are independent, and applying Szeg{\"o} Theorem to (\ref{eq:Eg1}), we can write
\begin{equation}\label{eq:Eg2}
\begin{split}
\E [g_1(\Lambda_i)]
& = \int_z \int_y \underbrace{\left(\int_x g_2(x, y, z) f_{X|Y,Z}(x|y,z) \: dx \right)}_{g_3(y,z)} \:\: f_{Y}(y) \:\: f_{Z}(z)  \: dy \: dz\\
& = \int_z \left(\int_y g_3(y,z) f_{Y}(y)\: dy\right) \:\: f_{Z} (z)  \: dz\\
& = \int_z \left(\frac{1}{2\pi} \int_{\lambda=0}^{2\pi} g_3\left(f_{r,i}(\lambda),z\right) \; d\lambda \right) \:\: f_{Z} (z)  \: dz \mbox{ , by Szeg{\"o} Theorem (\ref{eq:Szego})}\\
& = \frac{1}{2\pi} \int_{\lambda=0}^{2\pi} \left(\int_z g_3\left(f_{r,i}(\lambda),z\right) \:\: f_{Z} (z)  \: dz \right) \; d\lambda  \\
& =  \frac{1}{(2\pi)^2} \int_{\lambda=0}^{2\pi} \int_{\nu=0}^{2\pi} g_3\left(f_{r,i}(\lambda), f_{t,i+1}(\nu) \right) \; d\lambda \:d\nu \mbox{ , by Szeg{\"o} Theorem (\ref{eq:Szego})}.
\end{split}
\end{equation}

\vspace*{2mm}

\noindent\textbf{Equal power allocation over optimal precoding directions:}
We further assume equal power allocation over the optimal directions, i.e. the singular values of $\bP_i$ are chosen to be all equal: $\bLambda_{P_i}=\alpha_i \bI_{k_i}$, where $\alpha_i$ is real positive and chosen to respect the power constraint (\ref{eq:PowConstraints}). Equal power allocation may not be the optimal power allocation scheme, but it is considered in this example for simplicity.

Using the power constraint expression for general correlation models (\ref{eq:PowConsPi}) in Appendix \ref{ap:ProofTxDirec} and considering precoding matrices $\bP_i = \bU_{r,i}^H  (\alpha_i \bI_{k_i}) \bU_{t,i+1}$ with optimal singular vectors as in \emph{Theorem \ref{th:txDirections}} and equal singular values $\alpha_i$, we can show by induction on $i$ that the coefficients $\alpha_i$ respecting the power constraints for any correlation model are given by
\begin{equation}\label{eq:GenAlphai}
\begin{split}
\alpha_0 & = \sqrt{\mathcal{P}_0}\\
\alpha_i & = \sqrt{\frac{\mathcal{P}_i}{a_i \mathcal{P}_{i-1}} \frac{\tr(\bLambda_{r,i-1})}{\tr(\bLambda_{r,i})} \frac{k_i}{\tr(\bLambda_{t,i}\bLambda_{r,i-1})} } \qquad \forall i \in \{1,\ldots,N-1\}\\
\alpha_N & = 1.
\end{split}
\end{equation}
Applying the exponential correlation model to (\ref{eq:GenAlphai}) and making the dimensions of the system grow large, it can be shown that in the asymptotic regime, the $\alpha_i$ respecting the power constraint for the exponentially correlated system converge to the same value (\ref{eq:UncorrAlphai}) as for the uncorrelated system.

Then  $X =\Lambda_{P_i}^2= \alpha_i^2$ is independent from $Y$ and $Z$, thus $f_{X|Y,Z}(x|y,z)=f_X(x)=\delta(x-\alpha_i^2)$.
Consequently,
\begin{equation}
g_3(y,z)=\int_x g_2(x, y, z) \delta(x-\alpha_i^2) \: dx = g_2(\alpha_i^2,y,z)
\end{equation}
and (\ref{eq:Eg2}) becomes
\begin{equation}\label{eq:Egxyz}
\E [g_1(\Lambda_i)] = \frac{1}{(2\pi)^2} \int_{\lambda=0}^{2\pi} \int_{\nu=0}^{2\pi} g_2\left(\alpha_i^2,\frac{1-r_{r,i}^2}{|1-r_{r,i}e^{j\lambda}|^2}, \frac{1-r_{t,i+1}^2}{|1-r_{t,i+1}e^{j\nu}|^2} \right) \; d\lambda \:d\nu.
\end{equation}

\vspace*{2mm}

\noindent\textbf{Asymptotic Mutual Information}:
Using (\ref{eq:Egxyz}) in (\ref{eq:hh1}) with $g_2(x,y,z)=\log\left(1+\eta \frac{a_{i+1}}{\rho_i} h_i^N x y z \right)$  gives the expression of the asymptotic mutual information
\begin{equation}\label{eq:ExpI}
\begin{split}
\bbI_{\infty}=&
\sum_{i=0}^N \frac{\rho_i}{\rho_0(2\pi)^2}
 \int_{\lambda=0}^{2\pi} \int_{\nu=0}^{2\pi}
\log\left(1+ h_i^N \frac{\eta a_{i+1} \alpha_i^2 (1-r_{r,i}^2) (1-r_{t,i+1}^2)  }{\rho_i |1-r_{r,i}e^{j\lambda}|^2 |1-r_{t,i+1}e^{j\nu}|^2} \right) \; d\lambda \:d\nu - N\frac{\log e}{\rho_0}\eta\prod_{i=0}^N h_i
 \end{split}
\end{equation}
where $h_0,h_1,\ldots,h_N$ are the solutions of the following system of $N+1$ equations, obtained by using (\ref{eq:Egxyz}) in (\ref{eq:hh2}) with $g_2(x,y,z)=\frac{h_i^N\Lambda_i x y z}{ \frac{\rho_i}{a_{i+1}} + \eta h_i^N x y z}$
{\small
\begin{equation}\label{eq:Exphi}
\begin{split}
& \prod_{j=0}^N h_j=  \frac{\rho_i}{(2\pi)^2} \int_{\lambda=0}^{2\pi} \int_{\nu=0}^{2\pi}
\frac{h_i^N a_{i+1} \alpha_i^2 (1-r_{r,i}^2) (1-r_{t,i+1}^2)}{\rho_i |1-r_{r,i}e^{j\lambda}|^2 |1-r_{t,i+1}e^{j\nu}|^2 + \eta h_i^N a_{i+1} \alpha_i^2 (1-r_{r,i}^2) (1-r_{t,i+1}^2)} \; d\lambda \:d\nu\\
& \mbox{for } i=0,\ldots,N
\end{split}
\end{equation}}
(with the convention $r_{r,0}=r_{t,N+1}=0$).
Using the changes of variables
\begin{equation}
\begin{split}
t & =\tan\left(\frac{\lambda}{2}\right) \mbox{ , thus } \cos(\lambda)=\frac{1-t^2}{1+t^2} \quad \mbox{ and  } \quad d\lambda = \frac{2 du}{1+t^2}\\
u & =\tan\left(\frac{\nu}{2}\right)\mbox{ , thus } \cos(\nu)=\frac{1-u^2}{1+u^2} \quad \mbox{ and }  \quad d\nu = \frac{2 du}{1+u^2}
\end{split}
\end{equation}
and performing some  algebraic manipulations that are skipped for the sake of conciseness, (\ref{eq:ExpI}) and (\ref{eq:Exphi}) can be rewritten
{\small
\begin{equation}\label{eq:ExpI2}
\begin{split}
\bbI_{\infty}=&
\sum_{i=0}^N \frac{\rho_i}{\rho_0 \pi^2}
 \int_{t=-\infty}^{+\infty} \int_{u=-\infty}^{+\infty}
\log\left(1+ c_{r,i}{c_{t,i+1}} \frac{\eta h_i^N a_{i+1} \alpha_i^2}{\rho_i}
\frac{ (1+t^2)}{(c_{r,i}^2+t^2)} \frac{(1+u^2)}{(c_{t,i+1}^2+u^2)}
\right)
\; \frac{dt}{1+t^2} \: \frac{du}{1+u^2} - N\frac{\log e}{\rho_0}\eta\prod_{i=0}^N h_i
 \end{split}
\end{equation}}
where $h_0,h_1,\ldots,h_N$ are the solutions of the system of $N+1$ equations
\begin{equation}\label{eq:Exphi2}
\begin{split}
\prod_{j=0}^N h_j
& =\frac{2 }{\pi} \frac{ h_i^N a_{i+1} \alpha_i^2 }{  \sqrt{ c_{r,i} c_{t,i+1}  +  \frac{\eta h_i^N a_{i+1} \alpha_i^2}{\rho_i}  } \:\: \sqrt{ \frac{1}{c_{r,i} c_{t,i+1}}  +      \frac{\eta h_i^N a_{i+1} \alpha_i^2}{\rho_i}  }} \: K(m_i)
\end{split}
\end{equation}
and
\begin{equation}
\begin{split}
c_{r,i}&=\frac{1-r_{r,i}}{1+r_{r,i}}\\
c_{t,i+1}&=\frac{1-r_{t,i+1}}{1+r_{t,i+1}}\\
m_i &= 1-\frac{ \left( \frac{c_{t,i+1}}{c_{r,i}} +  \frac{\eta h_i^N a_{i+1} \alpha_i^2 }{\rho_i} \right)  \left(\frac{c_{r,i}}{c_{t,i+1}}  + \frac{\eta h_i^N a_{i+1} \alpha_i^2}{\rho_i}   \right)  }{ \left(\frac{1}{c_{r,i} c_{t,i+1}}  + \frac{\eta h_i^N a_{i+1} \alpha_i^2}{\rho_i}  \right)  \left(c_{r,i} c_{t,i+1} +  \frac{\eta h_i^N a_{i+1} \alpha_i^2}{\rho_i}  \right) } .
\end{split}
\end{equation}
Those expressions show that only a few relevant parameters affect the performance of this complex system: signal power $\mathcal{P}_i$, noise power $1/\eta$, pathloss $a_i$, number of hops $N$, ratio of the number of antennas $\rho_i$, and correlation ratios $c_{r,i}$ and $c_{t,i}$.

\section{Numerical Results}\label{sec:NumRes}

In this section, we present numerical results to validate \emph{Theorem \ref{th:Iasymptotic}} and to show that even with small $k_i,i=0,\ldots,N$, the behavior of the system is close to its behavior in the asymptotic regime, making \emph{Theorem \ref{th:Iasymptotic}} a useful tool for optimization of finite-size systems as well as large networks.

\subsection{Uncorrelated multi-hop MIMO}

The uncorrelated system described in Section \ref{sec:ComScen3} is first considered.

\begin{figure}[htbp]
  \centering
  \includegraphics[width=0.98\columnwidth]{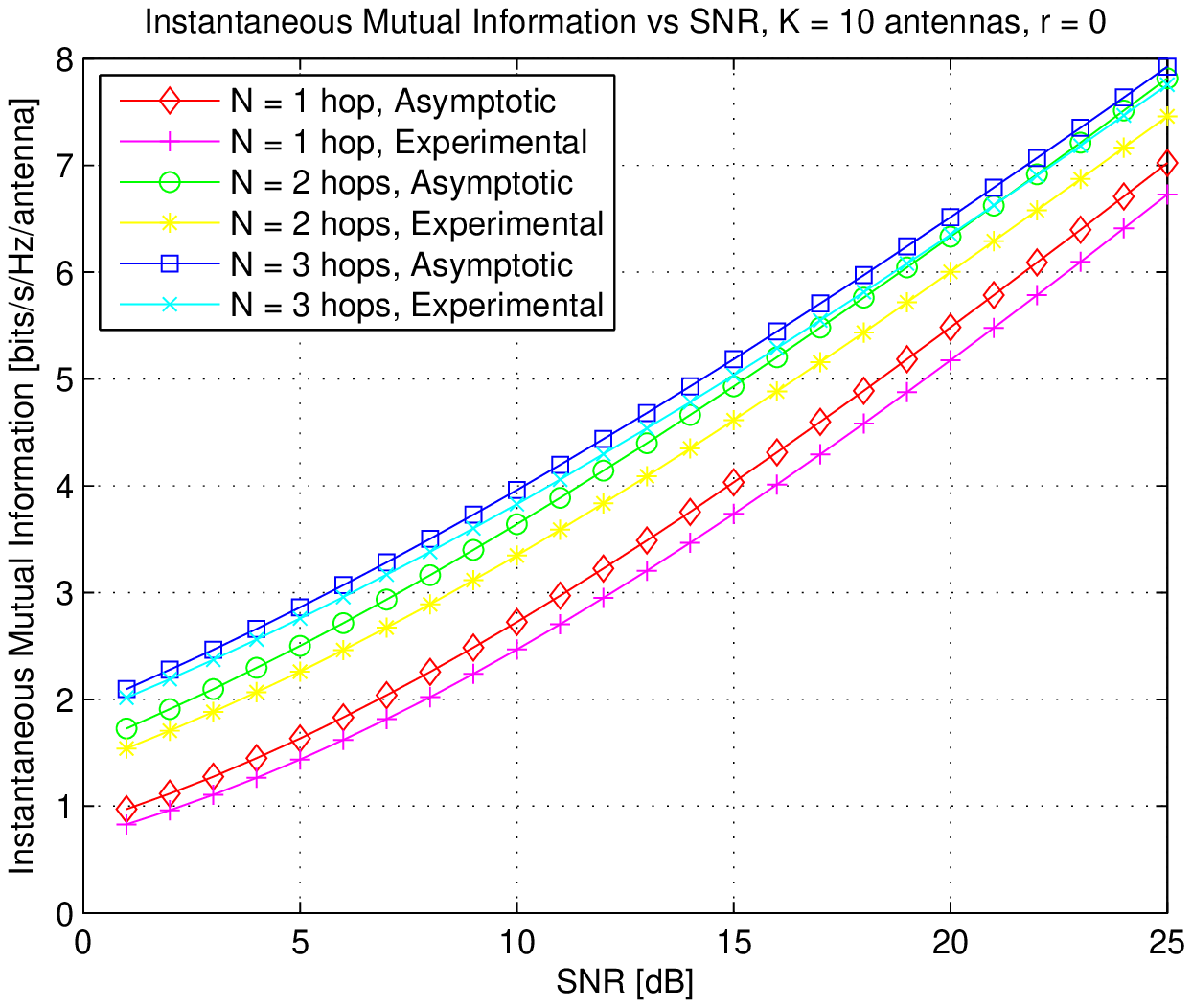}\\
  \caption[Instantaneous Mutual Information, 10 antennas]{Uncorrelated case: Asymptotic Mutual Information and Instantaneous Mutual Information versus SNR, with K = 10 antennas, for single-hop MIMO, 2 hops, and 3 hops}
  \label{fig:InstantMutInfo-K10}
\end{figure}

\begin{figure}[htbp]
  \centering
  \includegraphics[width=0.98\columnwidth]{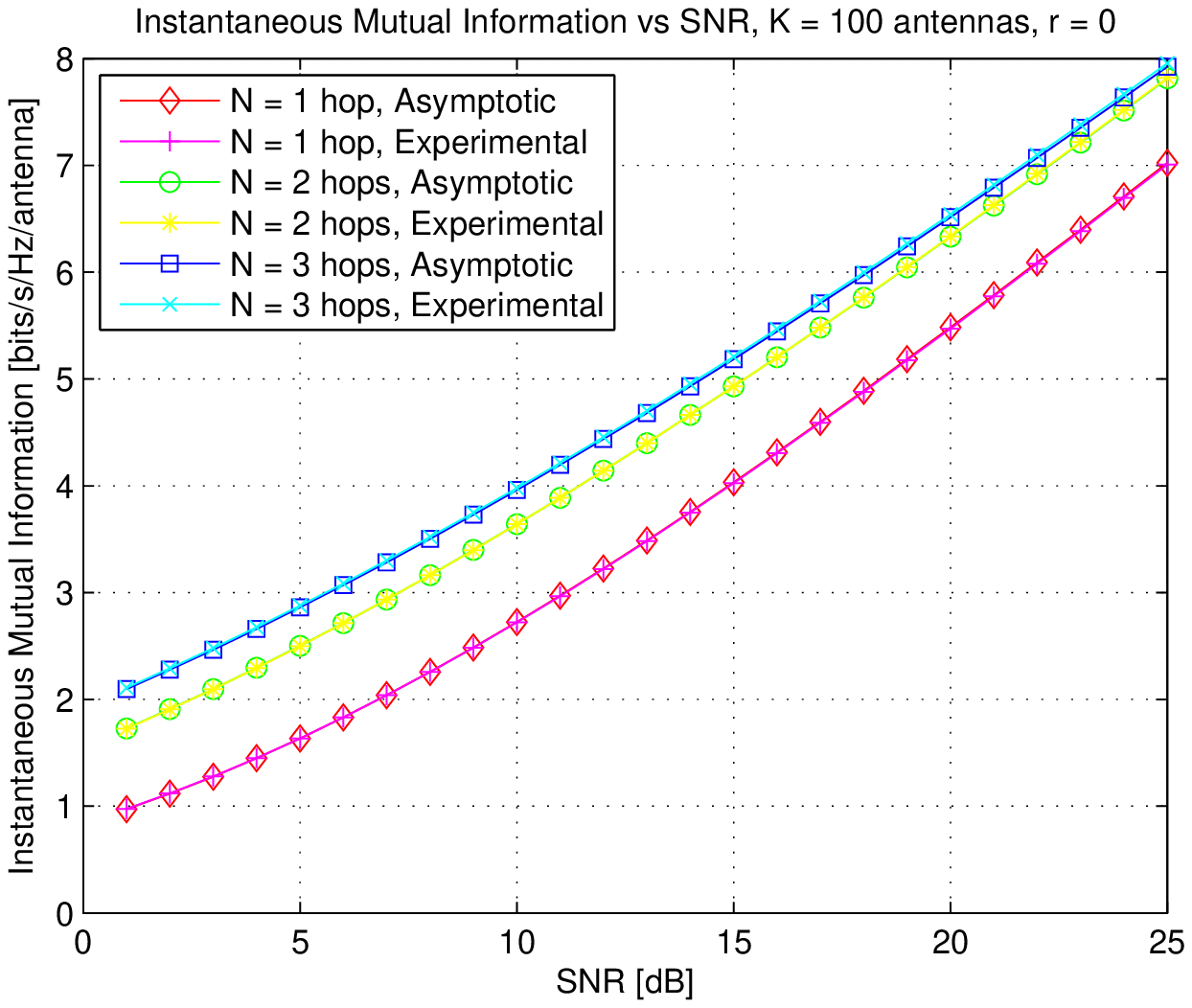}\\
  \caption[Instantaneous Mutual Information, 100 antennas]{Uncorrelated case: Asymptotic Mutual Information and Instantaneous Mutual Information versus SNR, with K = 100 antennas, for single-hop MIMO, 2 hops, and 3 hops}
  \label{fig:InstantMutInfo-K100}
\end{figure}

Fig. \ref{fig:InstantMutInfo-K10} plots the asymptotic mutual information from \emph{Theorem 1} as well as the instantaneous mutual information obtained for an arbitrary channel realization (shown as experimental curves in the figure). This example considers a system with $10$ antennas at source, destination and each relay level with one, two or three hops. 
Fig. \ref{fig:InstantMutInfo-K100} plots the same curves as in Fig. \ref{fig:InstantMutInfo-K10} for a system with $100$ antennas at each level. When increasing the number of hops $N$, the distance between source and destination $d$ is kept constant and $N-1$ relays are inserted between source and destination with equal spacing $d_i=d/N$ between each relaying level.
In both examples, whose main purpose is not to optimize the system, but to validate the asymptotic formula in \emph{Theorem \ref{th:Iasymptotic}}, matrices $P_i$ are taken proportional to the identity matrix to simulate equal power allocation. The channel correlation matrices are also equal to the identity matrix to mimic the uncorrelated channel. Moreover, the pathloss exponent $\beta=2$  is considered.
We would like to point out that the experimental curves for different channel realizations produced similar results. As such, the experimental curve corresponding to a single channel realization is shown for the sake of clarity and conciseness.

Fig. \ref{fig:InstantMutInfo-K100} shows the perfect match between the instantaneous mutual information for an arbitrary channel realization and the asymptotic mutual information, validating \emph{Theorem \ref{th:Iasymptotic}} for large network dimensions. On the other hand Fig. \ref{fig:InstantMutInfo-K10} shows that the instantaneous mutual information of a system with a small number of antennas behaves very closely to the asymptotic regime, justifying the usefulness of the asymptotic formula even when evaluating the end-to-end mutual information of a system with small size.

Finally, Fig. \ref{fig:InstantMutInfoVsK} plots the asymptotic mutual information for one, two, and three hops, as well as the value of the instantaneous mutual information for random channel realizations when the number of antennas at all levels increases. The concentration of the instantaneous mutual information values around the asymptotic limit when the system size increases shows the convergence of the instantaneous mutual information towards the asymptotic limit as the number of antennas grows large at all levels with the same rate.

\subsection{One-sided exponentially correlated multi-hop MIMO}

Based on the model discussed in Section \ref{sec:ComScen4}, the one-sided exponentially correlated system is considered in this section.
In the case of one-sided correlation, e.g. $r_{r,i}=0$ and $r_{t,i} \geq 0$ for all $i\in\{0,\ldots,N\}$, the asymptotic mutual information (\ref{eq:ExpI2}), (\ref{eq:Exphi2}) is reduced to {\small
\begin{equation}\label{eq:ExpI2}
\begin{split}
\bbI_{\infty}=&
\sum_{i=0}^N \frac{\rho_i}{\rho_0 \pi}
\int_{-\infty}^{+\infty}
\log\left(1+ {c_{t,i+1}} \frac{\eta h_i^N a_{i+1} \alpha_i^2}{\rho_i}
\frac{ (1+u^2)}{(c_{t,i+1}^2+u^2)}
\right)
 \: \frac{du}{1+u^2} - N\frac{\log e}{\rho_0}\eta\prod_{i=0}^N h_i
 \end{split}
\end{equation}
}
where $h_0,h_1,\ldots,h_N$ are the solutions of the system of $N+1$ equations
\begin{equation}\label{eq:Exphi2}
\begin{split}
\prod_{j=0}^N h_j
& = \frac{ h_i^N a_{i+1} \alpha_i^2 }{  \sqrt{  c_{t,i+1}  +  \frac{\eta h_i^N a_{i+1} \alpha_i^2}{\rho_i}  } \:\: \sqrt{ \frac{1}{ c_{t,i+1}}  +  \frac{\eta h_i^N a_{i+1} \alpha_i^2}{\rho_i}  }}.
\end{split}
\end{equation}
One-sided correlation was considered to avoid the involved computation of the elliptic integral $K(m_i)$ in the system of equations (\ref{eq:Exphi2}), and therefore to simplify simulations.

Fig. \ref{fig:CorrInstantMutInfo-K10} and \ref{fig:CorrInstantMutInfo-K100} plot the asymptotic mutual information for $10$ and $100$ antennas at each level respectively, and one, two or three hops, as well as the instantaneous mutual information obtained for an arbitrary channel realization (shown as experimental curves in the figure). As in the uncorrelated case, the perfect match of the experimental and asymptotic curves in Fig. \ref{fig:CorrInstantMutInfo-K100} with $100$ antennas validates the asymptotic formula in \emph{Theorem \ref{th:Iasymptotic} } in the presence of correlation. Fig. \ref{fig:CorrInstantMutInfo-K10} shows that even for a small number of antennas, the system behaves closely to the asymptotic regime in the correlated case.

\begin{figure}[htbp]
  \centering
  \includegraphics[width=0.98\columnwidth]{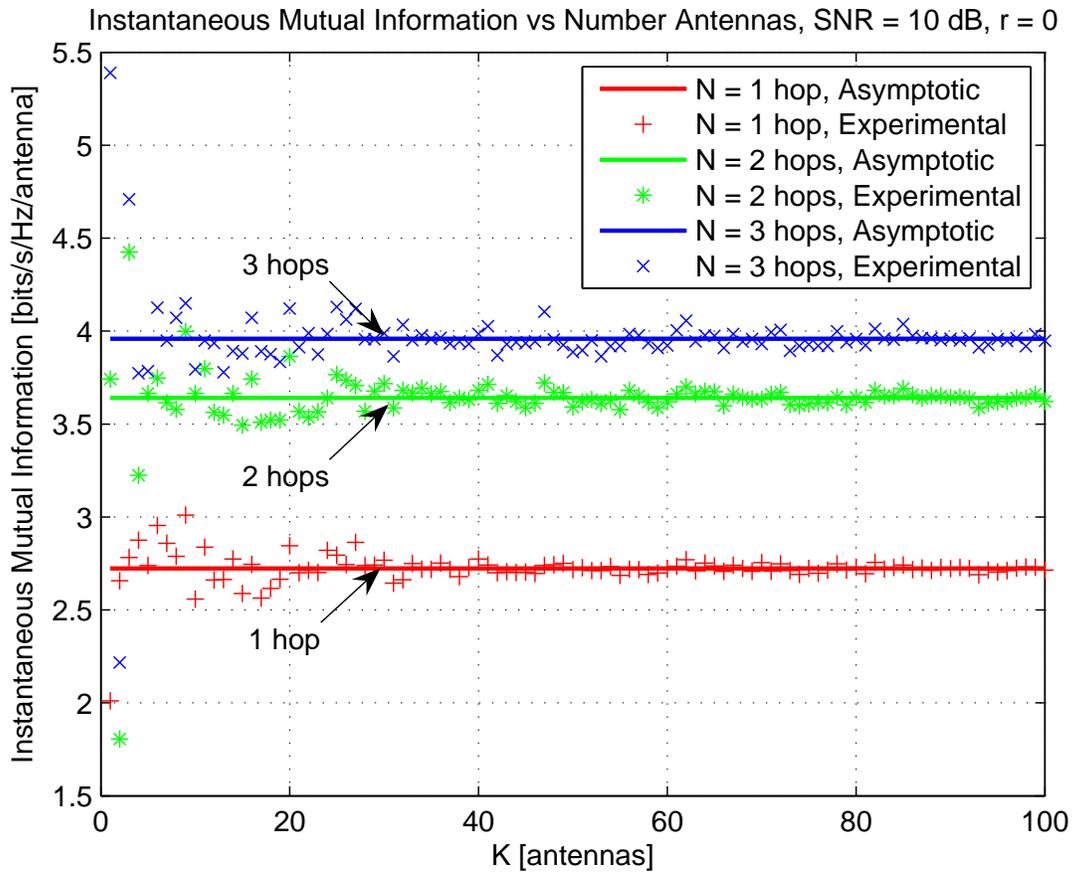}\\
  \caption[Instantaneous Mutual Information, SNR=10dB]{Uncorrelated case: Asymptotic Mutual Information and Instantaneous Mutual Information versus $K_N$, at SNR=10 dB, for single-hop MIMO, 2 hops, and 3 hops}
  \label{fig:InstantMutInfoVsK}
\end{figure}

\begin{figure}[htbp]
  \centering
  \includegraphics[width=0.98\columnwidth]{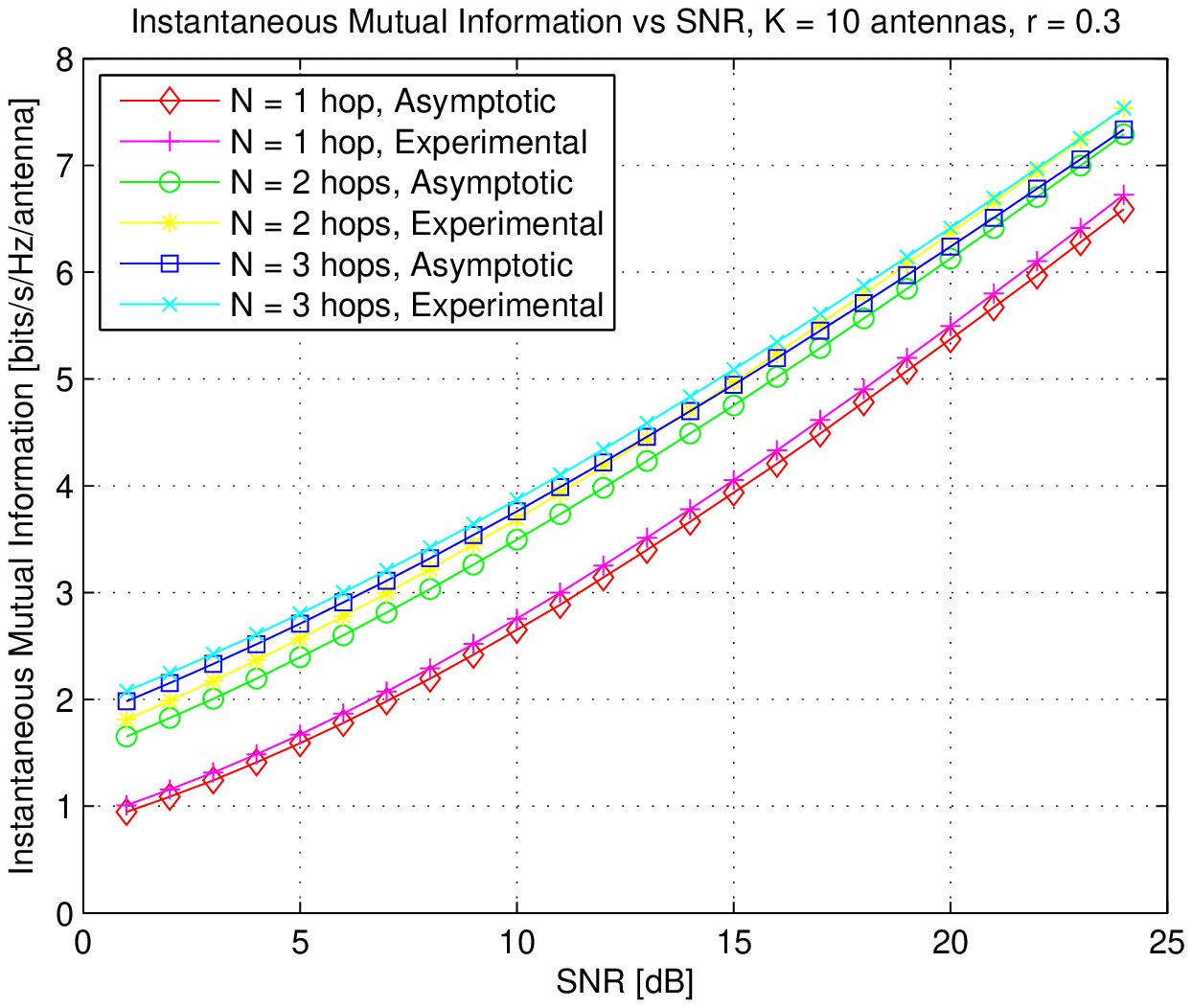}\\
  \caption[Instantaneous Mutual Information with Exponential Correlations, 10 antennas]{One-sided exponential correlation case: Asymptotic Mutual Information and Instantaneous Mutual Information versus SNR, with K = 10 antennas, r=0.3, for single-hop MIMO, 2 hops, and 3 hops}
  \label{fig:CorrInstantMutInfo-K10}
\end{figure}

\begin{figure}[htbp]
  \centering
  \includegraphics[width=0.98\columnwidth]{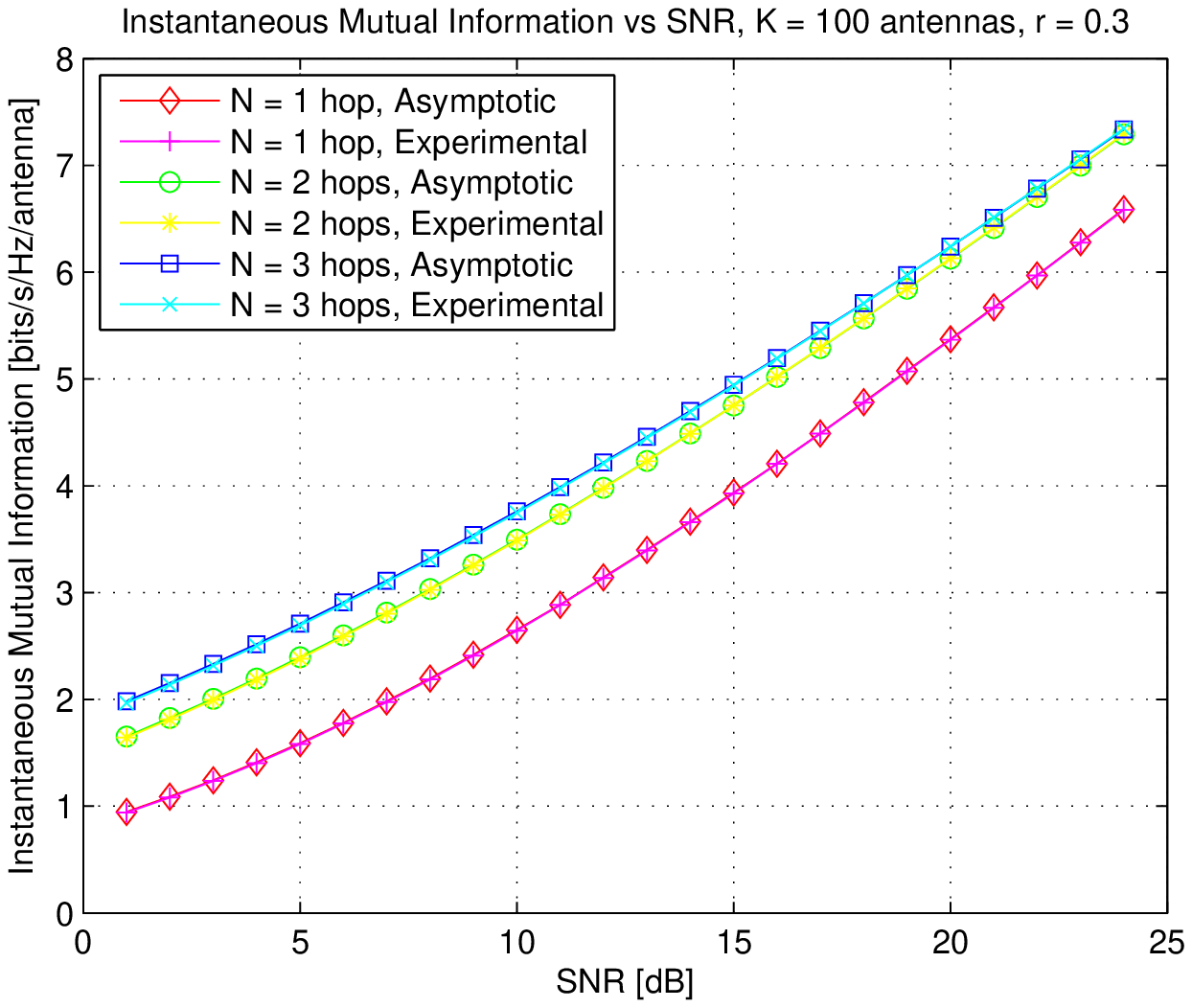}\\
  \caption[Instantaneous Mutual Information with Exponential Correlations, 100 antennas]{One-sided exponential correlation case: Asymptotic Mutual Information and Instantaneous Mutual Information versus SNR, with K = 100 antennas, r=0.3, for single-hop MIMO, 2 hops, and 3 hops}
  \label{fig:CorrInstantMutInfo-K100}
\end{figure}

\begin{figure}[htbp]
  \centering
  \includegraphics[width=0.98\columnwidth]{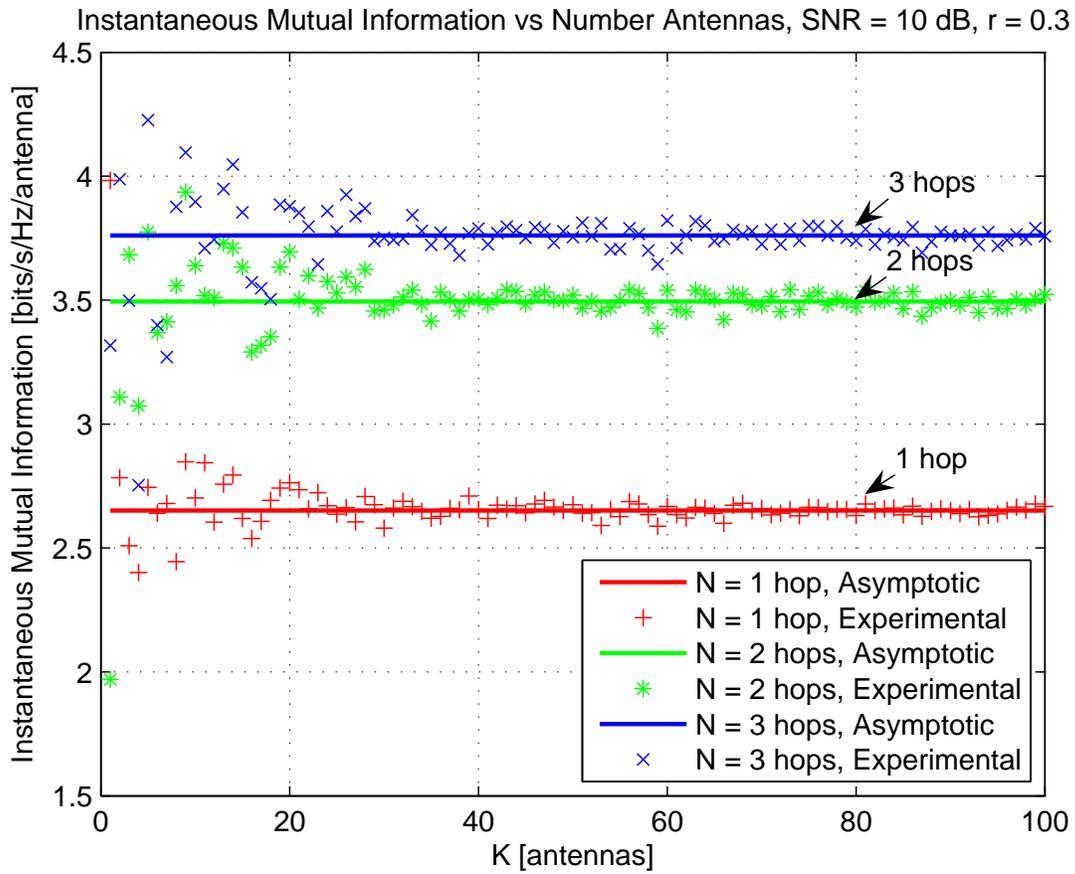}\\
  \caption[Instantaneous Mutual Information, SNR=10dB]{One-sided exponential correlation case: Asymptotic Mutual Information and Instantaneous Mutual Information versus $K_N$, at SNR=10 dB, r=0.3, for single-hop MIMO, 2 hops, and 3 hops}
  \label{fig:CorrInstantMutInfoVsK}
\end{figure}

Finally, Fig. \ref{fig:CorrInstantMutInfoVsK} plots the instantaneous mutual information for random channel realizations against the size of the system and shows its convergence towards the asymptotic mutual information when the number of antennas increases. Comparing Fig. \ref{fig:CorrInstantMutInfoVsK} to the corresponding Fig. \ref{fig:InstantMutInfoVsK} in the uncorrelated case, it appears that convergence towards the asymptotic limit is slower in the correlated case.

\section{Conclusion}\label{sec:Conclusion}

We studied a multi-hop MIMO relay network in the correlated fading environment, where relays at each level perform linear precoding on their received signal prior to retransmitting it to the next level.
Using free probability theory, a closed-form expression of the  instantaneous end-to-end mutual information was derived in the asymptotic regime where the number of antennas at all levels grows large.
The asymptotic instantaneous end-to-end mutual information turns out to be a deterministic quantity that depends only on channel statistics and not on particular channel realizations. Moreover, it also serves as the asymptotic value of the average end-to-end mutual information.
Simulation results verified that, even with a small number of antennas at each level, multi-hop systems  behave closely to the asymptotic regime. This observation makes the derived asymptotic mutual information a powerful tool to optimize the instantaneous mutual information of finite-size systems with only statistical knowledge of the channel.

We also showed that for any system size the left and right singular vectors of the optimal precoding matrices that maximize the average mutual information are aligned, at each level, with the eigenvectors of the transmit and receive correlation matrices of the forward and backward channels, respectively.
Thus, the singular vectors of the optimal precoding matrices can be determined with only local statistical channel knowledge at each level.

In the sequel, the analysis of the end-to-end mutual information in the asymptotic regime will first be extended to the case where noise impairs signal reception at each relaying level. Then, combining the expression of the asymptotic mutual information with the singular vectors of the optimal precoding matrices, future work will focus on optimizing the power allocation determined by the singular values of the precoding matrices.
Finally future research directions also include the analysis of the relay-clustering effect: given a total number of antennas $k_i$ at level $i$, instead of considering that the relaying level consists of a single relay equipped with many antennas ($k_i$), we can consider that a relaying level contains $n_i$ relays equipped with ($k_i/n_i$) antennas. Clustering has a direct impact on the structure of correlation matrices: when the $k_i$ antennas at level $i$ are distributed among several relays, correlation matrices become block-diagonal matrices, whose blocks represent the correlation between antennas at a relay, while antennas at different relays sufficiently separated in space are supposed uncorrelated. In the limit of a relaying level containing $k_i$ relays equipped with a single antenna, we fall back to the case of uncorrelated fading with correlation matrices equal to identity. The optimal size of clusters in correlated fading is expected to depend on the SNR regime.


\appendices

\section{Transforms and lemmas}\label{ap:MathTools}

Transforms and lemmas used in the proofs of \emph{Theorems \ref{th:Iasymptotic}} and \emph{\ref{th:txDirections}} are provided and proved in this appendix, while the proofs of \emph{Theorems \ref{th:Iasymptotic}} and \emph{\ref{th:txDirections}} are detailed in Appendices \ref{ap:ProofThAsymptoticMutInfo} and \ref{ap:ProofTxDirec}, respectively.

\subsection{Transforms}

Let $\bT$ be a square matrix of size $n$ with real eigenvalues $\lambda_{\bT}(1),\ldots,\lambda_{\bT}(n)$. The empirical eigenvalue distribution $F_{\bT}^n$ of $\bT$ is defined by
\begin{equation}
F_{\bT}^n(x)\triangleq \frac{1}{n}\sum_{i=1}^n u(x-\lambda_{\bT}(i)).
\end{equation}
We define the following transformations \cite{Mueller-2002}
\begin{eqnarray}
 \mbox{\emph{Stieltjes transform}:}& G_{\bT}(s) & \triangleq  \int \frac{1}{\lambda - s} dF_{\bT}(\lambda) \label{eq:StieltjesTrans}\\
  & \Upsilon_{\bT}(s) & \triangleq  \int\frac{s\lambda}{1-s\lambda}dF_{\bT}(\lambda) \label{ups} \\
 \mbox{\emph{S-transform}: } & S_{\bT}(z)& \triangleq  \frac{z+1}{z}\Upsilon_{\bT}^{-1}(z)  \label{Strans}
\end{eqnarray}
where $\Upsilon^{-1}(\Upsilon(s))=s$.

\subsection{Lemmas}

We present here the lemmas used in the proofs of \emph{Theorems \ref{th:Iasymptotic}} and \emph{\ref{th:txDirections}}. \emph{Lemmas \ref{lem:Mueller}},  \emph{\ref{lem:Mueller2}}, \emph{\ref{lem:Compacity}} and \emph{\ref{lem:traceProd}} are proved in Appendix \ref{sec:LemmaProofs}, while \emph{Lemmas \ref{lem:Hiai}}, \emph{\ref{lem:Szego}}, and \emph{\ref{lem:MarshallOlkin}} are taken from \cite{Hiai-Petz-2000}, \cite{Gray-2006}, and \cite{Marshall-Olkin-1979} respectively.

\bigskip

\begin{lemma}\label{lem:Mueller}
  Consider an $n \times p$ matrix $\bA$ and a $p \times n$ matrix $\bB$, such that their product $\bA\bB$ has non-negative real eigenvalues. Denote $\xi = \frac{p}{n}$. Then
    \begin{equation}\label{ssh}
    S_{\bA\bB}(z)=\frac{z+1}{z+\xi}S_{\bB\bA}\left(\frac{z}{\xi}\right).
    \end{equation}
\end{lemma}

\medskip

Note that \emph{Lemma \ref{lem:Mueller}} is a more general form of the results derived in \cite[Eq. (1.2)]{Silverstein-1995}, \cite[Eq. (15)]{Mueller-2002}.

\bigskip

\begin{lemma}[{\cite[Prop. 4.4.9 and 4.4.11]{Hiai-Petz-2000}}]\label{lem:Hiai}
For $n \in \mathds{N}$, let $p(n) \in \mathds{N}$ be such that $\frac{p(n)}{n} \rightarrow \xi$ as $n \rightarrow \infty$. Let
\begin{itemize}
\item $\bTheta(n)$ be a $ p(n) \times n$ complex Gaussian random matrix with i.i.d. elements with variance $\frac{1}{n}$.
\item $\bA(n)$ be a $n \times n$ constant matrix such that $\sup_n \|\bA(n) \| < + \infty$ and $(\bA(n),\bA(n)^H)$ has the limit eigenvalue distribution $\mu$.
\item $\bB(n)$ be a $p(n) \times p(n)$  Hermitian random matrix, independent from $\bTheta(n)$, with an empirical eigenvalue distribution converging almost surely to a compactly supported probability measure $\nu$.
\end{itemize}
Then, as $n \rightarrow \infty$,
\begin{itemize}
\item the empirical eigenvalue distribution of $\bTheta(n)^H \bB(n) \bTheta(n)$ converges almost surely to the compound free Poisson distribution $\pi_{\nu,\xi}$ \cite{Hiai-Petz-2000}
\item the family $(\{\bTheta(n)^H \bB(n) \bTheta(n)\},\{\bA(n),\bA(n)^H\})$ is asymptotically free almost everywhere.
\end{itemize}
Thus the limiting eigenvalue distribution of  $\bTheta(n)\bB(n)\bTheta(n)^H \bA(n)\bA(n)^H$ is the free convolution  $\pi_{\nu,\xi} \boxtimes \mu$ and its \emph{S-transform} is
    \begin{equation}\label{freeness}
        S_{\bTheta \bB \bTheta^H \bA\bA^H}(z)=S_{\bTheta\bB\bTheta^H}(z)S_{\bA\bA^H}(z).
    \end{equation}
\end{lemma}
Note that if the elements of $\bTheta(n)$ had variance $\frac{1}{p(n)}$ instead of $\frac{1}{n}$, $(\{\bTheta(n)^H \bB(n) \bTheta(n)\},\{\bA(n),\bA(n)^H\})$ would still be  asymptotically free almost everywhere, and consequently, Equation (\ref{freeness}) would still hold.

\bigskip

\begin{lemma}\label{lem:Mueller2}
    Consider an $n \times p$ matrix $\bA$ with zero-mean i.i.d. entries with variance $\frac{a}{p}$. Assume that the dimensions go to infinity while $\frac{n}{p}\rightarrow\zeta$, then
    \begin{equation}\label{sbazi}
    \begin{split}
    S_{\bA\bA^H}(z) & =\frac{1}{a}\:\: \frac{1}{(1+\zeta z)}\\
    S_{\bA^H \bA}(z) & = \frac{1}{a} \:\: \frac{1}{(z + \zeta)}.
    \end{split}
    \end{equation}
\end{lemma}

\bigskip

\begin{lemma}[{\cite[Theorem H.1.h]{Marshall-Olkin-1979}}]\label{lem:MarshallOlkin}
Let $\bA$ and $\bB$ be two positive semi-definite hermitian matrices of size $n\times n$. Let $\lambda_{\bA}(i)$  and $\lambda_{\bB}(i)$ be their decreasingly-ordered eigenvalues respectively. 
Then the following inequality holds:
\begin{equation}
  \sum_{i=1}^n \lambda_{\bA}(i) \lambda_{\bB}(n-i+1) \:\: \leq \:\: \tr(\bA \bB) = \sum_{i=1}^n \lambda_{\bA\bB}(i) \:\:\leq \:\: \sum_{i=1}^n \lambda_{\bA}(i) \lambda_{\bB}(i).
\end{equation}
\end{lemma}

\bigskip

\begin{lemma}\label{lem:Compacity}
    For $i\in\{1,\ldots,N\}$, let $\bA_i$ be a $n_i \times n_{i-1}$ random matrix. Assume that
    \begin{itemize}
      \item $\bA_1, \ldots, \bA_N$ are mutually independent,
      \item $n_i$  goes to infinity while $\frac{n_i}{n_{i-1}}\rightarrow\zeta_i$,
      \item as $n_i$ goes to infinity, the eigenvalue distribution of $\bA_i\bA_i^H$ converges almost surely in distribution to a compactly supported measure $\nu_i$,
      \item as $n_1,\ldots,n_N$ go to infinity, the eigenvalue distribution of $(\bigotimes_{i=N}^1 \bA_i)( \bigotimes_{i=N}^1 \bA_i)^H$ converges almost surely in distribution to a measure $\mu_N$.
    \end{itemize}

    Then $\mu_N$ is compactly supported.
\end{lemma}

\bigskip

\begin{lemma}[{\cite[Theorem 9]{Gray-2006}}]\label{lem:Szego}

Let $\bT_n$ be a sequence of Wiener-class Toeplitz matrices, characterized by the function $f(\lambda)$ with essential infimum $m_f$ and essential supremum $M_f$. Let $\lambda_{\bT_n}(1),\ldots,\lambda_{\bT_n}(n)$ be the eigenvalues of $\bT_n$ and $s$ be any positive integer. Then
\begin{equation}
\lim_{n\rightarrow \infty} \frac{1}{n}\sum_{k=1}^{n} \lambda_{\bT_n}^s(k)=\frac{1}{2\pi}\int_0^{2\pi}f(\lambda)^s d\lambda.
\end{equation}
Furthermore, if $f(\lambda)$ is real, or equivalently, the matrices $\bT_n$ are all Hermitian, then for any function $g(\cdot)$ continuous on $[m_f,M_f]$
\begin{equation}
\lim_{n\rightarrow \infty} \frac{1}{n}\sum_{k=1}^{n} g(\lambda_{\bT_n}(k))=\frac{1}{2\pi}\int_0^{2\pi} g(f(\lambda)) d\lambda.
\end{equation}
\end{lemma}


\begin{lemma}\label{lem:traceProd}
For $i \geq 1$, given a set of deterministic matrices $\{\bA_k \}_{k\in\{0,\dots,i\}}$ and a set of independent random matrices $\{\bTheta_k \}_{k\in\{1,\dots,i\}}$, with i.i.d. zero-mean  gaussian elements with variance $\sigma_k^2$,
\begin{equation}
\tr \left( \; \E \left[ \; \bigotimes_{k=i}^1 \{ \bA_k \bTheta_k\} \bA_0 \bA_0^H \bigotimes_{k=1}^i  \{ \bTheta_k^H \bA_k^H \} \; \right] \; \right) = \tr(\bA_0 \bA_0^H) \prod_{k=1}^i \sigma_k^2 \tr(\bA_k \bA_k^H).
\end{equation}
\end{lemma}

\bigskip


\subsection{Proofs of Lemmas}\label{sec:LemmaProofs}

The proofs of \emph{Lemmas \ref{lem:Mueller}}, \emph{\ref{lem:Mueller2}}, \emph{\ref{lem:Compacity}} and \emph{\ref{lem:traceProd}} are given hereafter.

\noindent \textbf{Proof of \emph{Lemma \ref{lem:Mueller}}}

Given two complex matrices $\bA$ of size $m \times n$,  and $\bB$ of size $n \times m$, their products $\bA\bB$ and $\bB\bA$ have the same $k$ non-zero eigenvalues $\lambda_{\bA\bB}(1),\ldots,\lambda_{\bA\bB}(k)$ with the same respective multiplicities $m_1,\ldots,m_k$.   However the multiplicities $m_0$ and $m_0'$ of the $0$-eigenvalues of $\bA\bB$ and $\bB\bA$ respectively, are related as follows:
\begin{equation}\label{eq:ZeroEV}
m_0 + n = m_0' + m.
\end{equation}

Assuming that $\bA\bB$, and therefore $\bB\bA$,  has 
real eigenvalues, we show hereafter that (\ref{ssh}) holds.

The empirical eigenvalue distributions of $\bA\bB$ and $\bB\bA$ are defined by
\begin{equation}
\begin{split}
F_{\bA\bB}^m(\lambda) &=\frac{m_0}{m} u(\lambda) + \frac{1}{m} \sum_{i=1}^{k} m_i u(\lambda-\lambda_{\bA\bB}(i))\\
F_{\bB\bA}^n(\lambda) &=\frac{m'_0}{n} u(\lambda) + \frac{1}{n} \sum_{i=1}^{k} m_i u(\lambda-\lambda_{\bA\bB}(i)).
\end{split}
\end{equation}
Using (\ref{eq:ZeroEV}), we get
\begin{equation}\label{eq:Frelation}
F_{\bA\bB}^m(\lambda) = \frac{n}{m} F_{\bB\bA}^n(\lambda) + \left( 1 - \frac{n}{m} \right) u(\lambda).
\end{equation}
From (\ref{eq:Frelation}), it is direct to show that
\begin{equation}\label{eq:Grelation}
G_{\bA\bB}(z) = \frac{n}{m} G_{\bB\bA}(z) - \left( 1 - \frac{n}{m} \right) \frac{1}{z}.
\end{equation}

As $\Upsilon(s) = -1 -\frac{1}{s}G(\frac{1}{s})$, from (\ref{eq:Grelation}), we obtain
\begin{equation}\label{eq:Upsrelation}
\Upsilon_{\bA\bB}(s) = \frac{n}{m} \Upsilon_{\bB\bA}(s).
\end{equation}
Finally, using $\{z=\Upsilon_{\bA\bB}(s) = \frac{n}{m} \Upsilon_{\bB\bA}(s)\} \Leftrightarrow \{\Upsilon_{\bA\bB}^{-1}(z)=s=\Upsilon^{-1}_{\bB\bA}\left(\frac{z}{n/m}\right) \}$
and  the definition of the \emph{S-transform} $S(z) \triangleq  \frac{z+1}{z}\Upsilon^{-1}(z)$ yields the desired result
\begin{equation}\label{eq:Srelation}
S_{\bA\bB}(z) = \frac{z+1}{z+\frac{n}{m}} S_{\bB\bA}\left(\frac{z}{n/m}\right).
\end{equation}
This concludes the proof of \emph{Lemma \ref{lem:Mueller}}.

\hfill $\blacksquare$

\bigskip


\noindent \textbf{Proof of \emph{Lemma \ref{lem:Mueller2}}}

Consider an $n \times p$ matrix $\bA$ with zero-mean i.i.d. entries with variance $\frac{a}{p}$. Let $\bX=\frac{1}{\sqrt{a}}\bA$  denote the normalized version of $\bA$ with zero-mean i.i.d. entries of variance $\frac{1}{p}$ and define $\bY=a \bI_n$ and $\bZ=\bX \bX^H \bY = \bA\bA^H$.
It is direct to show that $S_{\bY}(z)=\frac{1}{a}$. Using the latter result along with {\cite[Theorem 1]{Mueller-2002}}, we obtain
\begin{equation}
\begin{split}
S_{\bX\bX^H}(z) & = \frac{1}{(1+\zeta z)}  \\
S_{\bA\bA^H}(z) & = S_{\bZ}(z)=S_{\bX\bX^H}(z) S_{\bY}(z)= \frac{1}{(1+\zeta z)} \:\: \frac{1}{a}.
\end{split}
\end{equation}
Applying \emph{Lemma \ref{lem:Mueller}} to $S_{\bA^H \bA}(z)$ yields
\begin{equation}
S_{\bA^H \bA}(z) = \frac{z+1}{z+\zeta} S_{\bA\bA^H}\left(\frac{z}{\zeta}\right) = \frac{1}{a} \:\: \frac{1}{(z + \zeta)}.
\end{equation}
This completes the proof of \emph{Lemma \ref{lem:Mueller2}}.

\hfill $\blacksquare$

\bigskip

\noindent \textbf{Proof of \emph{Lemma \ref{lem:Compacity}}}

The proof of \emph{Lemma \ref{lem:Compacity}} is done by induction on $N$.
For $N=1$, \emph{Lemma \ref{lem:Compacity}} obviously holds.
Assuming that \emph{Lemma \ref{lem:Compacity}} holds for $N$, we now show  that it also holds for $N+1$.

We first recall that the eigenvalues of Gramian matrices $\bA\bA^H$ are non-negative. Thus the support of $\mu_{N+1}$ is lower-bounded by $0$, and we are left with showing that it is also upper-bounded.

Denoting $\bB_{N} = (\bigotimes_{i=N}^1 \bA_i)( \bigotimes_{i=N}^1 \bA_i)^H$, we can write
\begin{equation}
\bB_{N+1}= \bA_{N+1} \bB_N \bA_{N+1}^H.
\end{equation}

For a matrix $\bA$, let $\lambda_{\bA,\max}$ denote its largest eigenvalue.
The largest eigenvalue of $\bB_{N+1}$ is given by
\begin{equation}\label{eq:eigenvalueUpperbound}
\begin{split}
\lambda_{\bB_{N+1},\max}
& = \max_{\bx} \:\: \frac{\bx^H  \: \bB_{N+1}  \: \bx}{\bx^H \bx}\\
& = \max_{\bx} \:\: \frac{\bx^H  \: \bA_{N+1} \bB_N \bA_{N+1}^H  \:  \bx}{\bx^H \bx}\\
& = \max_{\bx} \:\: \frac{\tr(\bB_N  \:  \bA_{N+1}^H \bx \bx^H \bA_{N+1} )}{\bx^H \bx}\\
& \leq \max_{\bx} \:\: \frac{\sum_{k=1}^{n_N} \lambda_{\bB_N}(k)  \:  \lambda_{\bA_{N+1}^H \bx \bx^H \bA_{N+1}}(k)}{\bx^H \bx} \mbox{ , by \emph{Lemma \ref{lem:MarshallOlkin}}}\\
& \leq \max_{\bx} \:\: \lambda_{\bB_N,\max} \:\: \frac{\sum_{k=1}^{n_N}  \lambda_{\bA_{N+1}^H \bx \bx^H \bA_{N+1}}(k)}{\bx^H \bx}\\
& = \lambda_{\bB_N,\max} \:\: \max_{\bx} \:\: \frac{\tr(\bA_{N+1}^H \bx \bx^H \bA_{N+1})}{\bx^H \bx}\\
& = \lambda_{\bB_N,\max} \:\: \max_{\bx} \:\: \frac{\bx^H \bA_{N+1} \bA_{N+1}^H \bx }{\bx^H \bx}\\
& = \lambda_{\bB_N,\max} \:\: \lambda_{\bA_{N+1} \bA_{N+1}^H,\max}.
\end{split}
\end{equation}
To simplify notations, we rename the random variables as follows:
\begin{equation}
X =\lambda_{\bB_{N+1},\max} \qquad Y =\lambda_{\bB_N,\max} \qquad  Z =\lambda_{\bA_{N+1} \bA_{N+1}^H,\max}.
\end{equation}
Then (\ref{eq:eigenvalueUpperbound}) can be rewritten
\begin{equation}\label{eq:XYZ}
X \leq Y Z.
\end{equation}
Let $a \geq 0$, by (\ref{eq:XYZ}) we have
\begin{equation}\label{eq:FxFyz}
F_X(a)=\Pr\{X<a\} \geq \Pr\{YZ<a\} = F_{YZ}(a)
\end{equation}
which still holds for the asymptotic distributions as  $n_1,\ldots,n_{N+1} \rightarrow \infty$, while $\frac{n_i}{n_{i-1}}\rightarrow\zeta_i$. Denoting the plane region $\mathcal{D}_a=\{x,y \geq 0 / xy < a \}$, we can write
\begin{equation}
\begin{split}
F_{YZ}(a)
&= \iint_{y,z \in \mathcal{D}_a} f_{Y,Z}(y,z)dy dz \\
&= \iint_{y,z \in \mathcal{D}_a} f_{Y}(y) f_Z(z)dy dz \mbox{ , by independence of } Y \mbox{ and } Z\\
&= \int_{y=0}^{+\infty} \left(\int_{z=0}^{a/y} f_Z(z) dz \right) f_{Y}(y) dy\\
&= \int_{y=0}^{+\infty} F_Z\left(\frac{a}{y}\right) f_{Y}(y) dy.
\end{split}
\end{equation}

By assumption, the distributions of $\bA_{N+1} \bA_{N+1}^H$ and $\bB_N$ converge almost surely to compactly supported measures. Thus, their largest eigenvalues are asymptotically upper-bounded and the support of the asymptotic distributions of $Y$ and $Z$ are upper-bounded, i.e.
\begin{equation}
\begin{split}
& \exists c_y \geq 0 \mbox{ such that } \forall y \geq c_y \mbox{ , } F_Y(y)=1 \qquad (f_Y(y)=0) \\
& \exists c_z \geq 0 \mbox{ such that } \forall  z \geq c_z  \mbox{ , } F_Z(z)=1 \qquad (f_Z(z)=0).
\end{split}
\end{equation}

Let $a \geq c_y \: c_z$, then for all $ 0 < y \leq c_y$, we have  $ \frac{a}{y}\geq \frac{a}{c_y} \geq c_z$ and $F_Z\left(\frac{a}{y}\right)=1$, as the dimensions go to infinity with constant rates. Therefore, in the asymptotic regime, we have
\begin{equation}\label{eq:Fyz1}
\begin{split}
F_{YZ}(a)
&= \int_{y=0}^{c_y} F_Z\left(\frac{a}{y}\right) f_{Y}(y) dy\\
&= \int_{y=0}^{c_y} 1  f_{Y}(y) dy = F_Y(c_y)=1.
\end{split}
\end{equation}
Combining (\ref{eq:FxFyz}) and (\ref{eq:Fyz1}), we get $F_X(a)=1$ for $a > c_y \: c_z$.
Thus, there exists a constant $c_x$ such that $0 \leq c_x \leq c_y \: c_z$ and $\forall x \geq c_x \mbox{ , } F_X(x)=1$, which means that the support of the asymptotic distribution of $X$ is upper-bounded.
As a consequence, the support of the asymptotic eigenvalue distribution of $\bB_{N+1}$ is also upper-bounded. Therefore, the support of $\mu_{N+1}$ is upper-bounded, which concludes the proof.

\hfill $\blacksquare$

\bigskip


\noindent \textbf{Proof of \emph{Lemma \ref{lem:traceProd}}}

The proof of \emph{Lemma \ref{lem:traceProd}} is done by induction.

  We first prove that \emph{Lemma \ref{lem:traceProd}} holds for $i=1$.
  To that purpose, we define the matrix $\bB=\bA_1 \bTheta_1 \bA_0 \bA_0^H \bTheta_1^H \bA_1^H$. Then
  \begin{equation}\label{eq:Prodn1}
  \tr(\E [ \bA_1 \bTheta_1 \bA_0 \bA_0^H \bTheta_1^H \bA_1^H ])= \tr(\E [\bB]) =  \sum_{j=1}^{k_1} \E [b_{jj}]
  \end{equation}

  The expectation of the $j^{th}$ diagonal element $b_{jj}$  of matrix $\bB$ is
  \begin{equation}\label{eq:Prodn1xx}
    \begin{split}
  \E [b_{jj}] & = \sum_{k,l,m,n,p} \E [a^{(1)}_{jk} \theta^{(1)}_{kl} a^{(0)}_{lm} a^{(0)\ast}_{nm} \theta^{(1)\ast}_{pn} a^{(1)\ast}_{jp} ] \\
                 & = \sum_{k,l,m}  |a^{(1)}_{jk}|^2 |a^{(0)}_{lm}|^2 \underbrace{\E [ |\theta^{(1)}_{kl}|^2]}_{\sigma_1^2} \\ 
                 & = \sigma_1^2 \sum_k |a^{(1)}_{jk}|^2  \sum_{l,m} |a^{(0)}_{lm}|^2.
                 \end{split}
  \end{equation}
where the second equality is due to the fact that $\E [ \theta^{(1)}_{kl} \theta^{(1)\ast}_{pn} ]= \sigma_1^2 \delta_{k,p} \delta_{l,n}$.
It follows from (\ref{eq:Prodn1}) and (\ref{eq:Prodn1xx}) that
  \begin{equation}
    \tr(\E [\bB]) = \sigma_1^2 \sum_{j,k} |a^{(1)}_{jk}|^2  \sum_{l,m} |a^{(0)}_{lm}|^2
     = \sigma_1^2 \tr(\bA_1 \bA_1^H) \tr(\bA_0 \bA_0^H)
  \end{equation}
 which shows that \emph{Lemma \ref{lem:traceProd}} holds for $i=1$.

  Now, assuming that \emph{Lemma \ref{lem:traceProd}} holds for $i-1$, we show it also holds for $i$. We define the matrix $\bB_i=\bigotimes_{k=i}^1 \{ \bA_k \bTheta_k\} \bA_0 \bA_0^H \bigotimes_{k=1}^i  \{ \bTheta_k^H \bA_k^H \}$.

 Then
  \begin{equation}\label{eq:Prodni}
  \begin{split}
  \tr(\E [\bB_i] ) & = \tr(\E [\bA_i \bTheta_i \bB_{i-1} \bTheta_i^H \bA_i^H])\\
    & = \sum_{j=1}^{k_1} \E [b^{(i)}_{jj}].
  \end{split}
  \end{equation}

The expectation of the $j^{th}$ diagonal element $b^{(i)}_{jj}$  of matrix $\bB_i$ is
  \begin{equation}
    \begin{split}
  \E [b^{(i)}_{jj}] & = \sum_{k,l,m,n} \E [a^{(i)}_{jk} \theta^{(i)}_{kl} b^{(i-1)}_{lm} \theta^{(i)\ast}_{nm} a^{(i)\ast}_{jn} ] \\
                 & = \sum_{k,l}  |a^{(i)}_{jk}|^2 \E[b^{(i-1)}_{ll}] \underbrace{\E[ |\theta^{(i)}_{kl}|^2]}_{\sigma_i^2}\\
                 & = \sigma_i^2 \sum_k |a^{(i)}_{jk}|^2  \sum_{l} \E[b^{(i-1)}_{ll}]
                 \end{split}
  \end{equation}
where the second equality is due to the independence of $\bTheta_i $ and $\bB_{i-1}$ and to the fact that $\E [\theta^{(i)}_{kn} \theta^{(i)\ast}_{lm}] = \sigma_i^2 \delta_{k,p} \delta_{l,n}$.
Thus (\ref{eq:Prodni}) becomes
  \begin{equation}
  \begin{split}
    \tr(\E[\bB_i]) & = \sigma_i^2 \sum_{j,k} |a^{(i)}_{jk}|^2  \sum_{l} \E[b^{(i-1)}_{ll}]
     = \sigma_i^2 \tr(\bA_i \bA_i^H) \tr(\E[\bB_{i-1}])\\
     & = \sigma_i^2 \tr(\bA_i \bA_i^H) \tr(\bA_0 \bA_0^H) \prod_{k=1}^{i-1} \sigma_k^2 \tr(\bA_k \bA_k^H) = \tr(\bA_0 \bA_0^H) \prod_{k=1}^i  \sigma_k^2 \tr(\bA_k \bA_k^H)
     \end{split}
  \end{equation}
 which shows that if \emph{Lemma \ref{lem:traceProd}} holds for $i-1$, then it holds for $i$.

  Therefore \emph{Lemma \ref{lem:traceProd}} holds for any $i\geq1$, which concludes the proof.
\hfill $\blacksquare$

\bigskip

\section{Proof of Theorem \ref{th:Iasymptotic}}\label{ap:ProofThAsymptoticMutInfo}

In this appendix, we first list the main steps of the proof of \emph{Theorem \ref{th:Iasymptotic}} and then present the detailed proof of each step. Note that the proof of \emph{Theorem \ref{th:Iasymptotic}} uses tools from the free probability theory introduced in Appendix \ref{ap:MathTools}. The proof of \emph{Theorem \ref{th:Iasymptotic}} consists of the following four steps.
\begin{enumerate}
\item Obtain $S_{\bG_N\bG_N^H}(z)$.
\item Use $S_{\bG_N\bG_N^H}(z)$ to find $\Upsilon_{\bG_N\bG_N^H}(z)$.
\item Use $\Upsilon_{\bG_N\bG_N^H}(z)$ to obtain $d\bbI/d\eta$.
\item Integrate $d\bbI/d\eta$ to obtain $\bbI$ itself.
\end{enumerate}

\begin{itemize}
 \item \textbf{First Step: obtain $S_{\bG_N\bG_N^H}(z)$ }
\end{itemize}

\begin{theorem}\label{th:SGG}
 As $k_i,i=0,\ldots,N$ go to infinity with the same rate, the S-transform of $\bG_N\bG_N^H$ is given by
\begin{equation}\label{theo1}
 S_{\bG_N\bG_N^H}(z)=S_{\bM_N^H\bM_N}(z)\prod_{i=1}^N\frac{\rho_{i-1}}{a_i}\:\: \frac{1}{(z+\rho_{i-1})} S_{\bM_{i-1}^H\bM_{i-1}}\left(\frac{z}{\rho_{i-1}}\right).
\end{equation}
\end{theorem}

\vspace{0.5 cm}

\begin{proof}
The proof is done by induction using \emph{Lemmas \ref{lem:Mueller}, \ref{lem:Mueller2}, \ref{lem:Hiai}}.
First, we prove (\ref{theo1}) for $N=1$.
Note that
\begin{equation}\label{theo1_2}
\bG_1\bG_1^H=\bM_1\bTheta_1\bM_0\bM_0^H\bTheta_1^H\bM_1^H
\end{equation}
therefore
\begin{eqnarray}\label{theo1N1}
 S_{\bG_1\bG_1^H}(z) =& S_{\bTheta_1\bM_0\bM_0^H\bTheta_1^H\bM_1^H\bM_1}(z) & \mbox{ , by \emph{Lemma \ref{lem:Mueller}}} \nonumber\\
=& S_{\bTheta_1\bM_0\bM_0^H\bTheta_1^H}(z)S_{\bM_1^H\bM_1}(z) & \mbox{ , by \emph{Lemma \ref{lem:Hiai}}}  \nonumber\\
=& \frac{z+1}{z+\frac{k_0}{k_1}}S_{\bM_0\bM_0^H\bTheta_1^H\bTheta_1}\left(\frac{z}{\frac{k_0}{k_1}}\right)S_{\bM_1^H\bM_1}(z) & \mbox{ , by \emph{Lemma \ref{lem:Mueller}}} \nonumber\\
=& \frac{z+1}{z+\frac{k_0}{k_1}}S_{\bM_0\bM_0^H}\left(\frac{z}{\frac{k_0}{k_1}}\right)S_{\bTheta_1^H\bTheta_1}\left(\frac{z}{\frac{k_0}{k_1}}\right)S_{\bM_1^H\bM_1}(z) & \mbox{ , by \emph{Lemma \ref{lem:Hiai}}} \nonumber\\
=& \frac{z+1}{z+\frac{k_0}{k_1}}S_{\bM_0\bM_0^H}\left(\frac{z}{\frac{k_0}{k_1}}\right) \frac{1}{a_1} \:\: \frac{1}{\frac{z}{\frac{k_0}{k_1}} + \frac{k_1}{k_0}}
S_{\bM_1^H\bM_1}(z) & \mbox{ , by \emph{Lemma \ref{lem:Mueller2}}} \nonumber\\
=& S_{\bM_1^H\bM_1}(z) \:\: \frac{\rho_0}{a_1} \:\: \frac{1}{z+\rho_0}S_{\bM_0^H\bM_0}\left(\frac{z}{\rho_0}\right) & \mbox{ , by \emph{Lemma \ref{lem:Mueller}}}.
\end{eqnarray}
Now, we need to prove that if (\ref{theo1}) holds for $N=q$, it also holds for $N=q+1$.
Note that
\begin{equation}\label{tho1q1}
\bG_{q+1}\bG_{q+1}^H
=\bM_{q+1}\bTheta_{q+1}\bM_q\bTheta_q\ldots\bM_1\bTheta_1\bM_0\bM_0^H\bTheta_1^H\bM_1^H\ldots\bTheta_q^H\bM_q^H\bTheta_{q+1}^H\bM_{q+1}^H.
\end{equation}
Therefore,
\begin{eqnarray}\label{tho1Nqplus1}
S_{\bG_{q+1}\bG_{q+1}^H}(z)\!\!&\!\!=\!\!&\!\!S_{\bM_{q+1}\ldots\bM_{q+1}^H}(z) \nonumber\\
\!\!&\!\!=\!\!&\!\! S_{\bTheta_{q+1}\bM_q\ldots\bM_q^H\bTheta_{q+1}^H\bM_{q+1}^H\bM_{q+1}}(z) \mbox{ , by \emph{Lemma \ref{lem:Mueller}}}.
\end{eqnarray}
The empirical eigenvalue distribution of Wishart matrices $\bTheta_i \bTheta_i^H $ converges almost surely to the Mar\v{c}enko-Pastur law whose support is compact. Moreover, by assumption, the empirical eigenvalue distribution of $\bM_i^H\bM_i$, $i = 0,\ldots,N+1$ converges to an asymptotic distribution with a compact support. Thus, by \emph{Lemma~\ref{lem:Compacity}}, the asymptotic eigenvalue distribution of $\bM_q\bTheta_q\ldots\bTheta_q^H\bM_q^H$ has a compact support. Therefore \emph{Lemma~\ref{lem:Hiai}} can be applied to (\ref{tho1Nqplus1})to show that
\begin{eqnarray}\label{tho1Nqplus1bis}
S_{\bG_{q+1}\bG_{q+1}^H}(z)
\!\!&\!\!=\!\!&\!\! S_{\bTheta_{q+1}\ldots\bTheta_{q+1}^H}(z)S_{\bM_{q+1}^H\bM_{q+1}}(z) \mbox{ , by \emph{Lemma \ref{lem:Hiai}}}\nonumber\\
\!\!&\!\!=\!\!&\!\! \frac{z+1}{z+\frac{k_q}{k_{q+1}}}S_{\bM_q\ldots\bM_q^H\bTheta_{q+1}^H\bTheta_{q+1}}\left(\frac{z}{\frac{k_q}{k_{q+1}}}\right)S_{\bM_{q+1}^H\bM_{q+1}}(z) \mbox{ , by \emph{Lemma \ref{lem:Mueller}}} \nonumber\\
\!\!&\!\!=\!\!&\!\! \frac{z+1}{z+\frac{k_q}{k_{q+1}}}S_{\bM_q\ldots\bM_q^H}\left(\frac{z}{\frac{k_q}{k_{q+1}}}\right)S_{\bTheta_{q+1}^H\bTheta_{q+1}}\left(\frac{z}{\frac{k_q}{k_{q+1}}}\right)S_{\bM_{q+1}^H\bM_{q+1}}(z) \mbox{ , by \emph{Lemma \ref{lem:Hiai}}} \nonumber  \\
\!\!&\!\!=\!\!&\!\! \frac{z+1}{z+\frac{k_q}{k_{q+1}}} \left(S_{\bM_q^H\bM_q}\left(\frac{z}{\frac{k_q}{k_{q+1}}}\right)\prod_{i=1}^q \frac{\frac{k_{i-1}}{k_q}}{a_i} \:\: \frac{1}{\frac{z}{\frac{k_q}{k_{q+1}}}+\frac{k_{i-1}}{k_q}} S_{\bM_{i-1}^H\bM_{i-1}}\left(\frac{\left(\frac{z}{\frac{k_q}{k_{q+1}}}\right)}{\frac{k_{i-1}}{k_q}}\right)\right) \times \nonumber\\
\!\!&\!\!\!\!&\!\! \frac{1}{a_{q+1}} \:\: \frac{1}{\frac{k_{q+1}}{k_q}+\frac{z}{\frac{k_q}{k_{q+1}}}}
S_{\bM_{q+1}^H\bM_{q+1}}(z) \mbox{ , by \emph{Lemma \ref{lem:Mueller2}}} \nonumber\\
\!\!&\!\!=\!\!&\!\! \frac{z+1}{z+\frac{k_q}{k_{q+1}}}   S_{\bM_{q+1}^H\bM_{q+1}}(z) \frac{\frac{k_q}{k_{q+1}}}{a_{q+1}} \:\: \frac{1}{z+1} S_{\bM_q^H\bM_q}\left(\frac{z}{\frac{k_q}{k_{q+1}}}\right)
\prod_{i=1}^q
\frac{\frac{k_{i-1}}{k_{q+1}}}{a_i}
\:\: \frac{1}{z+\frac{k_{i-1}}{k_{q+1}}} S_{\bM_{i-1}^H\bM_{i-1}}\left(\frac{z}{\frac{k_{i-1}}{k_{q+1}}}\right) \nonumber\\
\!\!&\!\!=\!\!&\!\! S_{\bM_{q+1}^H\bM_{q+1}}(z) \prod_{i=1}^{q+1} \frac{\frac{k_{i-1}}{k_{q+1}}}{a_i}
\:\: \frac{1}{z+\frac{k_{i-1}}{k_{q+1}}} S_{\bM_{i-1}^H\bM_{i-1}}\left(\frac{z}{\frac{k_{i-1}}{k_{q+1}}}\right)\nonumber\\
\!\!&\!\!=\!\!&\!\! S_{\bM_{q+1}^H\bM_{q+1}}(z)\prod_{i=1}^{q+1} \frac{\rho_{i-1}}{a_i}
\:\: \frac{1}{(z+\rho_{i-1}) } S_{\bM_{i-1}^H\bM_{i-1}}\left(\frac{z}{\rho_{i-1}}\right).
\end{eqnarray}
The proof is complete.
\end{proof}

\begin{itemize}
 \item \textbf{Second Step: use $S_{\bG_N\bG_N^H}(z)$ to find $\Upsilon_{\bG_N\bG_N^H}(z)$}
\end{itemize}

\begin{theorem}\label{th:UpsilonGG}
Let us define $a_{N+1}=1$.
We have
\begin{equation}\label{theo2}
 s\Upsilon_{\bG_N\bG_N^H}^N(s)=\prod_{i=0}^N \:\: \frac{\rho_{i}}{a_{i+1}} \:\: \Upsilon^{-1}_{M_i^HM_i}\left(\frac{\Upsilon_{\bG_N\bG_N^H(s)}}{\rho_i}\right).
\end{equation}
\end{theorem}

\vspace{0.5 cm}

\begin{proof}
From (\ref{theo1}) it follows that
\begin{equation}\label{theo21}
 \frac{z}{z+1}S_{\bG_N\bG_N^H}(z) =  \frac{z}{z+1} S_{\bM_N^H\bM_N}(z) \prod_{i=1}^N \:\: \frac{\rho_{i-1}}{a_i} \:\: \frac{1}{z+\rho_{i-1}} \:\: \frac{\frac{z}{\rho_{i-1}}+1}{\frac{z}{\rho_{i-1}}} \:\: \frac{\frac{z}{\rho_{i-1}}}{\frac{z}{\rho_{i-1}}+1} \:\: S_{\bM_{i-1}^H\bM_{i-1}}\left(\frac{z}{\rho_{i-1}}\right).
\end{equation}
Using (\ref{Strans}) in (\ref{theo21}), we obtain
\begin{equation}\label{theo22}
\Upsilon_{\bG_N\bG_N^H}^{-1}(z)=\frac{1}{z^N}\Upsilon^{-1}_{M_{N}^HM_{N}}(z)\prod_{i=1}^N \:\: \frac{\rho_{i-1}}{a_i} \:\: \Upsilon^{-1}_{M_{i-1}^HM_{i-1}}\left(\frac{z}{\rho_{i-1}}\right)
\end{equation}
or, equivalently,
\begin{equation}\label{theo23}
\Upsilon_{\bG_N\bG_N^H}^{-1}(z)=\frac{1}{z^{N}}\prod_{i=0}^N \:\: \frac{\rho_i}{a_{i+1}} \:\: \Upsilon^{-1}_{M_{i}^HM_{i}}\left(\frac{z}{\rho_i}\right).
\end{equation}
Substituting $z=\Upsilon_{\bG_N\bG_N^H}(s)$ in (\ref{theo23}), Equation (\ref{theo2}) follows. This completes the proof.
\end{proof}

\begin{itemize}
 \item \textbf{Third Step: use $\Upsilon_{\bG_N\bG_N^H}(z)$ to obtain $d\bbI/d\eta$}
\end{itemize}

\begin{theorem}\label{th:dIdeta}
In the asymptotic regime, as $k_0, k_1,\ldots, k_N$ go to infinity while $\frac{k_i}{k_N}\rightarrow\rho_i, i=0,\ldots, N$, the derivative of the instantaneous mutual information is given by
\begin{equation}\label{theo3}
\frac{d\bbI_{\infty}}{d\eta}=\frac{1}{\rho_0\ln2} \prod_{i=0}^N h_i
\end{equation}
where $h_0,h_1,\ldots,h_N$ are the solutions to the following set of $N+1$ equations
\begin{equation}\label{theo32}
\prod_{j=0}^N h_j=\rho_i\E\left[\frac{h_i^N\Lambda_i}{ \frac{\rho_i}{a_i+1} +\eta h_i^N\Lambda_i}\right]\qquad i=0,\ldots,N.
\end{equation}
The expectation in (\ref{theo32}) is over $\Lambda_i$ whose probability distribution function is given by $F_{\bM_i^H\bM_i}(\lambda)$ (convention: $a_{N+1}=1$). 
\end{theorem}

\vspace{0.5 cm}

\begin{proof}

First, we note that
\begin{eqnarray}\label{Iint}
\bbI&=&\frac{1}{k_0}\log\det(\bI+\eta\bG_N\bG_N^H)\nonumber\\
&=&\frac{1}{k_0} \sum_{i=1}^{k_N}\log(1+\eta\lambda_{\bG_N\bG_N^H}(i))\nonumber\\
&=&\frac{k_N}{k_0} \frac{1}{k_N}\sum_{i=1}^{k_N}\log(1+\eta\lambda_{\bG_N\bG_N^H}(i))\nonumber\\
&=&\frac{k_N}{k_0}\int\log(1+\eta\lambda)dF^{k_N}_{\bG_N\bG_N^H}(\lambda)\nonumber\\
&\stackrel{a.s.}{\rightarrow}&\frac{1}{\rho_0}\int\log(1+\eta\lambda)dF_{\bG_N\bG_N^H}(\lambda) \nonumber\\
&=&\frac{1}{\rho_0\ln 2}\int\ln(1+\eta\lambda)dF_{\bG_N\bG_N^H}(\lambda)
\end{eqnarray}
where $F^{k_N}_{\bG_N\bG_N^H}(\lambda)$ is the (non-asymptotic) empirical  eigenvalue distribution of $\bG_N\bG_N^H$, that converges almost-surely to the asymptotic empirical  eigenvalue distribution $F_{\bG_N\bG_N^H}$, whose support is compact. Indeed, the empirical eigenvalue distribution of Wishart matrices $\bTheta_i \bTheta_i^H $ converges almost surely to the Mar\v{c}enko-Pastur law whose support is compact, and by assumption, for $i \in \{0,\ldots,N+1\}$ the empirical eigenvalue distribution of $\bM_i^H\bM_i$ converges to an asymptotic distribution with a compact support. Therefore, according to \emph{Lemma \ref{lem:Compacity}}, the asymptotic eigenvalue distribution of $\bG_N\bG_N^H$ has a compact support. The $\log$ function is continuous, thus bounded on the compact support of the asymptotic eigenvalue distribution of $\bG_N\bG_N^H$. This enables the application of the bounded convergence theorem to obtain the almost-sure convergence in (\ref{Iint}).

It follows from (\ref{Iint}) that
\begin{eqnarray}\label{theo33}
 \frac{d\bbI_{\infty}}{d\eta}&=&\frac{1}{\rho_0 \ln 2}\int\frac{\lambda}{1+\eta\lambda}dF_{\bG_N\bG_N^H}(\lambda)\nonumber\\
&=&\frac{1}{-\rho_0\eta \ln 2}\int\frac{-\eta\lambda}{1-(-\eta)\lambda}dF_{\bG_N\bG_N^H}(\lambda)\nonumber\\
&=&\frac{1}{-\rho_0\eta \ln 2}\Upsilon_{\bG_N\bG_N^H}(-\eta).
\end{eqnarray}
Let us denote
\begin{eqnarray}
 t& = &\Upsilon_{\bG_N\bG_N^H}(-\eta)\label{theo34}\\
g_i& = &\Upsilon^{-1}_{\bM_i^H\bM_i}\left(\frac{t}{\rho_i}\right)\qquad i=0,\ldots,N \label{theo35}
\end{eqnarray}
and, for the sake of simplicity, let $\alpha=\rho_0\ln 2$. From (\ref{theo33}), we have
\begin{equation}\label{theo36}
 t=-\eta\alpha\frac{d\bbI_{\infty}}{d\eta}.
\end{equation}
Substituting $s=-\eta$ in (\ref{theo2}) and using (\ref{theo34}) and (\ref{theo35}), it follows that
\begin{equation}\label{theo37}
-\eta t^N = \prod_{i=0}^N \:\: \frac{\rho_i}{a_{i+1}} \:\: g_i.
\end{equation}
Finally, from (\ref{theo35}) and the very definition of  $\Upsilon$ in (\ref{ups}), we obtain
\begin{equation}\label{theo38}
 t=\rho_i\int\frac{g_i\lambda}{1-g_i\lambda}dF_{\bM_i^H\bM_i}(\lambda)\qquad i=0,\ldots,N.
\end{equation}
Substituting (\ref{theo36}) in (\ref{theo37}) and (\ref{theo38}) yields
\begin{equation}\label{gg1}
 (-\eta)^{N+1}\left(\alpha\frac{d\bbI}{d\eta}\right)^N=\prod_{i=0}^N \:\: \frac{\rho_i}{a_{i+1}} \:\: g_i
\end{equation}
and
\begin{equation}\label{gg2}
-\eta\left(\alpha\frac{d\bbI_{\infty}}{d\eta}\right)=\rho_i\int\frac{g_i\lambda}{1-g_i\lambda}dF_{\bM_i^H\bM_i}(\lambda)\qquad i=0,\ldots,N.
\end{equation}
Letting
\begin{equation}\label{gg3}
 h_i \: = \: \left(\frac{\rho_i}{a_{i+1}}\right)^{\frac{1}{N}} \: \left(\frac{g_i}{-\eta}\right)^{\frac{1}{N}}
\end{equation}
it follows from (\ref{gg1}) that
\begin{equation}\label{gg4}
\alpha\frac{d\bbI_{\infty}}{d\eta}=\prod_{i=0}^N h_i.
\end{equation}
Using (\ref{gg3}) and (\ref{gg4}) in (\ref{gg2}), we obtain
\begin{equation}\label{gg5}
-\eta \prod_{j=0}^N h_j=\rho_i\int\frac{-\eta h_i^N \: \frac{a_{i+1}}{\rho_i} \: \lambda}{1-(-\eta) h_i^N \frac{a_{i+1}}{\rho_i}\lambda}dF_{\bM_i^H\bM_i}(\lambda)\qquad i=0,\ldots,N
\end{equation}
or, equivalently,
\begin{eqnarray}\label{gg6}
\prod_{j=0}^N h_j&=&\rho_i\int\frac{h_i^N\lambda}{ \frac{\rho_i}{a_{i+1}} +\eta h_i^N\lambda}dF_{\bM_i^H\bM_i}(\lambda)\nonumber\\
&=& \rho_i\E\left[\frac{h_i^N\Lambda_i}{\frac{\rho_i}{a_{i+1}}  +\eta h_i^N\Lambda_i}\right]\qquad i=0,\ldots,N.
\end{eqnarray}
This, along with equation (\ref{gg4}), complete the proof.
\end{proof}

\begin{itemize}
 \item \textbf{Fourth Step: integrate $d\bbI/d\eta$ to obtain $\bbI$ itself}
\end{itemize}

The last step of the proof of \emph{Theorem \ref{th:Iasymptotic}} is accomplished by computing the derivative of $\bbI_{\infty}$ in (\ref{eq:hh1}) with respect to $\eta$ and showing that the derivative matches (\ref{theo3}). This shows that (\ref{eq:hh1}) is one primitive function of $\frac{d\bbI_{\infty}}{d\eta}$. Since primitive functions of $\frac{d\bbI_{\infty}}{d\eta}$ differ by a constant, the constant was chosen such that the mutual information (\ref{eq:hh1}) is zero when SNR $\eta$ goes to zero: $\lim_{\eta \rightarrow 0} \bbI_{\infty}(\eta) = 0$.

We now proceed with computing the derivative of $\bbI_{\infty}$. If (\ref{eq:hh1}) holds, then we have (recall $\alpha=\rho_0\ln 2$)
\begin{equation}\label{hh3}
\alpha\bbI_{\infty}=\sum_{i=0}^N\rho_i\E\left[\ln\left(1 + \frac{\eta a_{i+1}}{\rho_i} h_i^N\Lambda_i\right)\right]-N\eta\prod_{i=0}^N h_i.
\end{equation}
From (\ref{hh3}) we have
\begin{eqnarray}\label{hh4}
\alpha\frac{d\bbI_{\infty}}{d\eta}
\!\!&\!\!=\!\!&\!\! \sum_{i=0}^N \rho_i \E\left[\frac{ \Lambda_i \left(h_i^N+ N\eta  h_i^{N-1} h_i^{'}\right)}{ \frac{\rho_i}{a_{i+1}} (1+ \frac{\eta a_{i+1}}{\rho_i} h_i^N\Lambda_i)}\right] - N\prod_{i=0}^Nh_i-N\eta\left(\sum_{i=0}^Nh_i^{'}\prod_{\stackrel{\scriptstyle{j=0}}{j\neq i}}^N h_j\right)\nonumber\\
\!\!&\!\!=\!\!&\!\! \sum_{i=0}^N\rho_i \E\left[\frac{\Lambda_i h_i^N}{ \frac{\rho_i}{a_{i+1}} +\eta h_i^N\Lambda_i}\right]
+ N\eta  \sum_{i=0}^N \frac{h_i^{'}}{h_i}\rho_i \E\left[\frac{\Lambda_i h_i^N }{ \frac{\rho_i}{a_{i+1}} +\eta h_i^N\Lambda_i}\right]
- N\prod_{i=0}^Nh_i - N\eta\left(\sum_{i=0}^N \frac{h_i^{'}}{h_i}\prod_{j=0}^N h_j\right)\nonumber\\
\!\!&\!\!=\!\!&\!\! \sum_{i=0}^N\prod_{j=0}^N h_j+N\eta  \left(\sum_{i=0}^N \frac{h_i^{'}}{h_i}\prod_{j=0}^N h_j\right)-N\prod_{i=0}^Nh_i-N\eta\left(\sum_{i=0}^N \frac{h_i^{'}}{h_i}\prod_{j=0}^N h_j\right)\nonumber\\
\!\!&\!\!=\!\!&\!\! (N+1)\prod_{j=0}^N h_j-N\prod_{j=0}^N h_j\nonumber\\
\!\!&\!\!=\!\!&\!\!\prod_{j=0}^N h_j
\end{eqnarray}
where $h_i^{'}\triangleq\frac{dh_i}{d\eta}$ and the third line is due to (\ref{eq:hh2}). Equation (\ref{theo3}) immediately follows from (\ref{hh4}).
This completes the proof.

\hfill $\blacksquare$

\section{Proof of Theorem \ref{th:txDirections}}\label{ap:ProofTxDirec}

In this appendix, we provide the proof of \emph{Theorem \ref{th:txDirections}}.
The proof of this theorem  is based on \cite[Theorem H.1.h]{Marshall-Olkin-1979} that is reiterated in \emph{Lemma \ref{lem:MarshallOlkin}}. Note that, \cite[Theorem H.1.h]{Marshall-Olkin-1979} has been used before to characterize the source precoder maximizing the average mutual information of single-user \cite{Jafar-Goldsmith-2004} and multi-user \cite{Soysal-Ulukus-2007} single-hop MIMO systems with covariance knowledge at source, and to obtain the relay precoder maximizing the instantaneous mutual information of a two-hop MIMO system with full CSI at the relay \cite{TH07}. We extend the results of \cite{Jafar-Goldsmith-2004}, \cite{Soysal-Ulukus-2007},  \cite{TH07} to suit the MIMO multi-hop relaying system of our concern.

The proof consists of three following steps.
\begin{itemize}
  \item \textbf{Step 1: }
  Use the singular value decomposition (SVD)
  $\bU_i \bD_i \bV_i^H = \bLambda_{t,i+1}^{1/2} \bU_{t,i+1}^H \bP_i \bU_{r,i}\bLambda_{r,i}^{1/2}$
  and show that unitary matrices $\bU_i$ and $\bV_i$ impact the maximization of the average mutual information through the power constraints only, while diagonal matrices $\bD_i$ affect  both the mutual information expression and the power constraints.

  \item \textbf{Step 2: } Represent the power constraint expression as a function of $\bD_i, \bU_i, \bV_i$ and channel correlation matrices only.

  \item \textbf{Step 3: } Show that the directions minimizing the trace in the power constraint 
      are those given in \emph{Theorem \ref{th:txDirections}}, regardless of the singular values contained in $\bD_i$.
\end{itemize}

Before detailing each step, we recall that the maximum average mutual information is given by
\begin{equation}
\begin{split}
 \bbC & \triangleq \max_{\{\bP_i / \tr(\E[\bx_i \bx_i^H]) \leq k_i \mathcal{P}_i \}_{i\in\{0,\ldots,N-1\}}}
 \E\left[\log \det(\bI_{k_N}+ \eta \; \bG_N\bG_N^H)\right]
\end{split}
\end{equation}
and we define the conventions $a_0=1$, and $\bC_{r,0}=\bI_{k_0}$. Note that the latter implies that $\bU_{r,0} =\bI_{k_0}$ and $ \bLambda_{r,0} = \bI_{k_0}$.

\vspace{0.5 cm}

\begin{itemize}
 \item \textbf{Step 1: clarify how the average mutual information depends on the transmit directions and the transmit powers}
\end{itemize}

For $i\in\{1,\ldots,N\}$ we define
\begin{equation}\label{eq:Zi}
\bTheta'_i=\bU_{r,i}^H \bTheta_i \bU_{t,i}
\end{equation}
Since $\bTheta_i$ is zero-mean i.i.d. complex Gaussian, thus bi-unitarily invariant, and $\bU_{r,i}$ and $\bU_{t,i}$ are unitary matrices, $\bTheta'_i$ has the same distribution as $\bTheta_i$.

For $i\in\{0,\ldots,N-1\}$, we consider the following SVD 
\begin{equation}\label{eq:UiDi}
\bU_i \bD_i \bV_i^H = \bLambda_{t,i+1}^{1/2} \bU_{t,i+1}^H \bP_i \bU_{r,i}\bLambda_{r,i}^{1/2} 
\end{equation}
where $\bU_i$, $\bV_i$ are unitary matrices, $\bD_i$ is a real diagonal matrix with non-negative diagonal elements in the non-increasing order of amplitude.

We now rewrite the average mutual information as a function of matrices $\bU_i$, $\bV_i$ and $\bD_i$, in order to take the maximization in (\ref{eq:Capa}) over $\bU_i$, $\bV_i$ and $\bD_i$ instead of $\bP_i$.
Using (\ref{eq:Zi}) and (\ref{eq:UiDi}) the average mutual information $\I$ can be expressed in terms of matrices $\bTheta'_i$, $\bU_i$, $\bV_i$ and $\bD_i$ as
\begin{equation}
\begin{split}
\I
&\triangleq \E \left[\log \det (\bI_{k_N}+ \eta \; \bG_N\bG_N^H)\right]\\
       &= \E \left[ \log \det  ( \bI_{k_N}+ \eta \; \bU_{r,N} \bLambda_{r,N}^{1/2}  \; \bTheta'_N \; \bU_{N-1} \bD_{N-1} \bV_{N-1}^H \; \bTheta'_{N-1}  \ldots \bU_1  D_1 \bV_1^H \;  \bTheta'_1 \; \bU_0 \bD_0 \bV_0^H \right.\\
        & \;\;\;\;\;\;\;\;\;\;\;\;\;\;\;\;\;\;\;\;\;\;\;\;\;\;\;\;\;\;\;\;\;
         \left. \bV_0 \bD_0^H \bU_0^H \; \bTheta_1^{'H} \; \bV_1 D_1^H \bU_1^H \ldots  \bTheta_{N-1}^{'H} \; \bV_{N-1} \bD_{N-1}^H \bU_{N-1}^H \; \bTheta_N^{'H} \bLambda_{r,N}^{1/2} \bU_{r,N}^H )\right]
  \end{split}
\end{equation}

$\bTheta'_i$ being zero-mean i.i.d. complex Gaussian, multiplying it by unitary matrices does not change its distribution. Therefore, $\bTheta''_i=\bV_{i}^H \bTheta'_i\bU_{i-1}$ has the same distribution as $\bTheta'_i$ and the average mutual information can be rewritten
\begin{equation}\label{eq:mutInfAvg}
\begin{split}
\I
&= \E \left[ \log \det  ( \bI_{k_N}+ \eta \;  \bLambda_{r,N}^{1/2} \bTheta''_N  \bD_{N-1} \bTheta''_{N-1} \ldots \bD_1 \bTheta''_1  \bD_0 \bD_0^H \bTheta_1^{''H} \bD_1^H \ldots  \bTheta_{N-1}^{''H} \bD_{N-1}^H \bTheta_N^{''H} \bLambda_{r,N}^{1/2} ) \right]\\
&= \E \left[ \log \det  ( \bI_{k_N}+ \eta \; \bLambda_{r,N}^{1/2}
            \;\;  \bigotimes_{i=N}^1 \{ \bTheta''_i  \bD_{i-1}   \}
            \;\;  \bigotimes_{i=1}^{N}  \{  \bD_{i-1}^H \bTheta_i^{''H}\}
            \;\;  \bLambda_{r,N}^{1/2} ) \right].
\end{split}
\end{equation}
Therefore, the maximum average mutual information can then be represented as
\begin{equation}\label{eq:CapaUiDi}
\begin{split}
 \bbC & = \!\!\! \max_{
               \begin{array}{c}
              \bD_i,\bU_i,\bV_i \\
              \tr(\E[\bx_i \bx_i^H]) \leq k_i \mathcal{P}_i \\
              \forall i\in\{0,\ldots,N-1\}
              \end{array}}
\!\!\!\!\!\!\!\!\!\!\!\!\!
\E \left[ \log \det  ( \bI_{k_N}+ \eta \; \bLambda_{r,N}^{1/2}
            \;\;  \bigotimes_{i=N}^1 \{ \bTheta''_i  \bD_{i-1}   \}
            \;\;  \bigotimes_{i=1}^{N}  \{  \bD_{i-1}^H \bTheta_i^{''H}\}
            \;\;  \bLambda_{r,N}^{1/2} ) \right].
\end{split}
\end{equation}
Expression (\ref{eq:mutInfAvg}) shows that the average mutual information $\I$ does not depend on the matrices $\bU_i$ and $\bV_i$, which determine the transmit directions at source and relays, but only depends on the singular values contained in matrices $\bD_i$.
Nevertheless, as shown by (\ref{eq:CapaUiDi}), the maximum average mutual information $\bbC$ depends on the matrices $\bU_i,\bV_i$---and thus on the transmit directions--- through the power constraints.

\vspace{0.5 cm}

\begin{itemize}
  \item \textbf{Step 2: give the expression of the power constraints in function of $\bD_i, \bU_i, \bV_i$ and channel correlation matrices}
\end{itemize}

We show hereunder that the average power of transmitted signal $\bx_i$ at $i$-th relaying level is given by
\begin{equation}\label{eq:PowCons}
\tr(\E[\bx_i \bx_i^H])= a_i \tr(\bP_i \bC_{r,i} \bP_i^H) \;\;\; \prod_{k=0}^{i-1} \frac{a_k}{k_k} \tr( \bC_{t,k+1} \bP_k \bC_{r,k} \bP_k^H ).
\end{equation}

\vspace{0.5 cm}

\begin{proof}
The average power of transmitted signal $\bx_i$ can be written as
\begin{equation*}
\tr(\E [ \bx_i \bx_i^H ] )= \tr(\E [\bigotimes_{k=i}^1 \{ \bA_k \bTheta_k\} \bA_0 \bA_0^H \bigotimes_{k=1}^i  \{ \bTheta_k^H \bA_k^H \} ])
\end{equation*}
with
\begin{equation}\label{eq:Ak}
    \begin{split}
    \bA_i & = \bP_i \bC_{r,i}^{1/2}\\
    \bA_k & = \bM_k = \bC_{t,k+1}^{1/2} \bP_k \bC_{r,k}^{1/2} \mbox{ , } \forall k \in \{0,\ldots,i-1\}\\
    \sigma_k^2 & = \frac{a_k}{k_{k-1}}
    \end{split}
\end{equation}
Applying \emph{Lemma \ref{lem:traceProd}} to $\tr({\rm E} \{ \bx_i \bx_i^H \} )$ yields
\begin{equation}
\begin{split}
\tr(\E[\bx_i \bx_i^H])
& =  \tr( \bC_{t,1} \bP_0 \bC_{r,0} \bP_0^H)   \;\;\; \prod_{k=1}^{i-1} \frac{a_k}{k_{k-1}} \tr( \bC_{t,k+1} \bP_k \bC_{r,k} \bP_k^H ) \;\;\; \frac{a_i}{k_{i-1}} \tr(\bP_i \bC_{r,i} \bP_i^H) \\
&= a_i \tr(\bP_i \bC_{r,i} \bP_i^H) \;\;\; \prod_{k=0}^{i-1} \frac{a_k}{k_k} \tr( \bC_{t,k+1} \bP_k \bC_{r,k} \bP_k^H )
\end{split}
\end{equation}
which concludes the proof.
\end{proof}

Using (\ref{eq:PowCons}) in the power constraints (\ref{eq:PowConstraints}), those constraints can be rewritten as a product of trace-factors:
\begin{equation}\label{eq:PowConsPi}
\begin{split}
& \tr(\bP_0 \bP_0^H) \leq k_0 \mathcal{P}_0 \\
& a_i \tr(\bP_i \bC_{r,i} \bP_i^H) \;\; \prod_{k=0}^{i-1} \frac{a_k}{k_k} \tr( \bC_{t,k+1} \bP_k \bC_{r,k} \bP_k^H )  \leq k_i \mathcal{P}_i \mbox{ , }  \forall i\in\{1,\ldots,N-1\}.
\end{split}
\end{equation}

In order to express (\ref{eq:PowConsPi}) in function of matrices $\bU_i$, $\bV_i$ and $\bD_i$, we first rewrite (\ref{eq:UiDi}) as
\begin{equation}\label{eq:Pi}
\bP_i = \bU_{t,i+1} \bLambda_{t,i+1}^{-1/2} \bU_i \bD_i \bV_i^H \bLambda_{r,i}^{-1/2} \bU_{r,i}^H
\end{equation}
and use (\ref{eq:Pi}) in (\ref{eq:PowConsPi}) to obtain
\begin{equation}
\begin{split}
\tr(\bP_i \bC_{r,i} \bP_i^H) & =\tr(\bU_{t,i+1} \bLambda_{t,i+1}^{-1/2} \bU_i \bD_i \bV_i^H \bLambda_{r,i}^{-1/2} \bU_{r,i}^H \;\;\; \bU_{r,i}\bLambda_{r,i}\bU_{r,i}^H  \;\;\;
\bU_{r,i} \bLambda_{r,i}^{-1/2} \bV_i \bD_i^H \bU_i^H \bLambda_{t,i+1}^{-1/2} \bU_{t,i+1}^H) \\
                            & =\tr( \bLambda_{t,i+1}^{-1} \bU_i \bD_i^2 \bU_i^H )\\
\tr( \bC_{t,k+1} \bP_k \bC_{r,k} \bP_k^H )
&= \tr(\bD_k \bD_k^H)\\
& = \tr(\bD_k^2)
\end{split}
\end{equation}
where $\bD_i^2=\bD_i \bD_i^H$ is a real diagonal matrix with non-negative diagonal elements in non-increasing order. 
This leads to the following expression of the power constraints in function of $\bU_i, \bD_i$
\begin{equation}\label{eq:PowConsUiDi}
\begin{split}
& \tr(\bLambda_{t,1}^{-1} \bU_0 \bD_0^2 \bU_0^H) \leq k_0 \mathcal{P}_0 \\
& a_i \tr( \bLambda_{t,i+1}^{-1} \bU_i \bD_i^2 \bU_i^H ) \leq \frac{ k_i \mathcal{P}_i}{\prod_{k=0}^{i-1} \frac{a_k}{k_{k}} \tr( \bD_k ^2 )}
\mbox{ , }  \forall i\in\{2,\ldots,N-1\}.
\end{split}
\end{equation}
It was shown in Step 1 that matrices $\bV_i$ do not have an impact on the expression of the average mutual information $\I$ (\ref{eq:mutInfAvg}), and surprisingly (\ref{eq:PowConsUiDi}) now shows that matrices $\bV_i$ do not have an impact on the power constraints either. In fact, as can be observed from (\ref{eq:PowConsUiDi}), the power constraints depend only on matrices $\bU_i$ and $\bD_i$.
It should also be noticed that matrix $\bU_i$ has an impact on the power constraint of the $i$-th relay only.

\vspace{0.5 cm}

\begin{itemize}
 \item \textbf{Step 3: give the optimal transmit directions}
\end{itemize}

To determine the optimal directions of transmission at source, we apply \emph{Lemma \ref{lem:MarshallOlkin}} to the source power constraint (\ref{eq:PowConsUiDi}) $\tr(\bLambda_{t,1}^{-1} \bU_0 \bD_0^2 \bU_0^H) \leq k_0 \mathcal{P}_0$, and conclude that for all choices of diagonal elements of $\bD_0^2$, the matrix $\bU_0$ that minimizes the trace $\tr(\bLambda_{t,1}^{-1} \bU_0 \bD_0^2 \bU_0^H)$ is $\bU_0=I_{k_0}$.
Therefore, the source precoder becomes
\begin{equation}
\begin{split}
\bP_0
&= \bU_{t,1} \bLambda_{t,1}^{-1/2} \bD_0  \bV_0^H \bLambda_{r,0}^{-1/2} \bU_{r,0}^H = \bU_{t,1} \bLambda_{t,1}^{-1/2} \bD_0 \bV_0^H\\
&= \bU_{t,1} \bLambda_{P_0} \bV_0^H.
\end{split}
\end{equation}
This recalls the known result (\ref{xs2}) in the single-hop MIMO case, where the optimal precoding covariance matrix at source was shown \cite{Jafar-Goldsmith-2004,Soysal-Ulukus-2007} to be
\begin{equation}
\bQ^\star \triangleq \E[\bx_0 \bx_0^H] = \bP_0\bP_0^H =\bU_{t,1}\bLambda_{\bQ^\star}\bU_{t,1}^H.
\end{equation}

Similarly, to determine the optimal direction of transmission at $i$-th relaying level, we apply \emph{Lemma~ \ref{lem:MarshallOlkin}} to the $i$-th power constraint:
for all choices of diagonal elements of $\bD_i^2$,
the matrix $\bU_i$ that minimizes the trace $\tr( \bLambda_{t,i+1}^{-1} \bU_i \bD_i^2 \bU_i^H)$  is $\bU_i=I_{k_i}$.
This leads to the precoding matrix at level $i$
\begin{equation}
\bP_i = \bU_{t,i+1} \bLambda_{t,i+1}^{-1/2} \bD_i \bV_i^H \bLambda_{r,i}^{-1/2} \bU_{r,i}^H.
\end{equation}

Now since matrices $\bV_i, i\in\{0,\ldots,N-1\}$ have an impact neither on the expression of the average mutual information nor on the power constraints, they can be chosen to be equal to identity: $\bV_i = \bI, i\in\{0,\ldots,N-1\}$ . This leads to the (non-unique but simple) optimal precoding matrices
\begin{equation}
\begin{split}
\bP_0 &= \bU_{t,1}  \bLambda_{P_0}\\
\bP_i &= \bU_{t,i+1}  \bLambda_{P_i} \bU_{r,i}^H
\end{split}
\end{equation}
with the diagonal matrices $\bLambda_{P_i} = \bLambda_{t,i+1}^{-1/2} \bD_i  \bLambda_{r,i}^{-1/2}$ containing the singular values of $\bP_i$.

This completes the proof of Theorem \ref{th:txDirections}.
\hfill $\blacksquare$

\section*{Acknowledgments}

The authors would like to thank B.H. Khan for his valuable help.


\bibliographystyle{./Biblio/IEEEtran}

\bibliography{./Biblio/IEEEabrv,./Biblio/bibLargeNet}


\end{document}

%% file: sty/macros.tex
\newtheorem{theorem}{Theorem}
\newtheorem{lemma}{Lemma}

\newfont{\mbb}{msbm10 scaled 1100}

\def\tr{{\rm tr}} 
\def\E{{\rm E}} 
\def\H{{\mathcal{H}}} 
\def\I{{\mathcal{I}}} 
\def\SNR{{\scriptstyle{\bf SNR}}} 

\def\bbI{{\mathcal{\boldsymbol I}}}
\def\bbC{{\mathcal{\boldsymbol C}}}
\def\bAs{{\bf A_s}}
\def\bAr{{\bf A_r}}
\def\bAd{{\bf A_d}}

\def\b0{{\bf 0}}

\def\bx{{\bf x}}
\def\by{{\bf y}}
\def\bz{{\bf z}}

\def\bA{{\bf A}}
\def\bB{{\bf B}}
\def\bC{{\bf C}}
\def\bD{{\bf D}}

\def\bG{{\bf G}}
\def\bH{{\bf H}}
\def\bI{{\bf I}}
\def\bM{{\bf M}}

\def\bP{{\bf P}}
\def\bQ{{\bf Q}}

\def\bT{{\bf T}}
\def\bU{{\bf U}}
\def\bV{{\bf V}}

\def\bX{{\bf X}}
\def\bY{{\bf Y}}
\def\bZ{{\bf Z}}

\def\bLambda{{\bf \Lambda}}

\def\bOmega{{\bf \Omega}}
\def\bTheta{{\bf \Theta}}